\documentclass[preprint,11pt,authoryear]{article}
\usepackage{amsmath}
\usepackage{amsfonts}
\usepackage{amssymb}
\usepackage{amsbsy}
\usepackage{amsthm}
\usepackage{geometry}
\usepackage{color,graphics,subfigure}
\usepackage{hyperref}
\usepackage{pifont}
\usepackage{lineno}
\usepackage{bm}
\usepackage{setspace}
\usepackage{cite}
\usepackage{graphicx}
\usepackage{latexsym,multicol}
\usepackage{tcolorbox}
\numberwithin{equation}{section}

\def\F{\mathbf{F}}

\def\S{\mathbf{S}}

\def \div{\mbox{div\hskip 1pt}}

\def \grad{\mbox{grad\hskip 1pt}}

\begin{document}
\title{Elephant trunk wrinkles:\\ A mathematical model of function and form}
\author{Yang Liu\footnote{Mathematical Institute, University of Oxford, Woodstock Road, Oxford, OX2 6GG, UK, Department of Mechanics, School of Mechanical Engineering, Tianjin University, Tianjin 300354, China}
\quad Alain Goriely\footnote{Mathematical Institute, University of Oxford, Woodstock Road, Oxford, OX2 6GG, UK}
\quad L. Angela Mihai\footnote{Corresponding author: L.A. Mihai, Email: \texttt{MihaiLA@cardiff.ac.uk}, School of Mathematics, Cardiff University, Senghennydd Road, Cardiff, CF24 4AG, UK}
}
\date{}
\maketitle
\begin{abstract}
\hrule\vskip 12pt
A notable feature of the elephant trunk is the pronounced wrinkling that enables its great flexibility. Here, we devise a general mathematical model that accounts for the characteristic skin wrinkles formed during morphogenesis in the elephant trunk. Using physically realistic parameters and operating within the theoretical framework of nonlinear morphoelasticity, we elucidate analytically and numerically the effect of skin thickness, relative stiffness, and differential growth on the physiological pattern of transverse wrinkles distributed along the trunk. We conclude that since the skin and muscle components have similar material properties, geometric parameters, such as curvature, play an important role. In particular, our model predicts that, in the proximal region close to the skull, where the curvature is lower, fewer wrinkles form and will form sooner than in the distal narrower region, where more wrinkles develop. Similarly, less wrinkling is found on the ventral side, which is flatter, compared to the dorsal side. In summary, the mechanical compatibility between the skin and the muscle enables them to grow seamlessly, while the wrinkled skin acts as a protective barrier that is both thicker and more flexible than the unwrinkled skin. 
\vskip 6pt
\hrule\vskip 6pt
\noindent{\bf Keywords:} hyperelastic solids, nonlinear deformation, instabilities, wrinkling, bilayer systems, mathematical modeling.
\end{abstract} 

\begin{quote}
\begin{flushright}
``Elephants have wrinkles, wrinkles, wrinkles,\\
Elephants have wrinkles, wrinkles everywhere.\\
On their trunks, on their ears, on their knees,\\
On their toes, no one knows, no one knows\\
Why-I-I-I-Ih!''-- Nursery Rhyme
\end{flushright}
\end{quote}


\section{Introduction}
The elephant trunk (Latin \emph{proboscis}, Ancient Greek $\pi\rho o\gamma o\sigma\kappa$\'{i}\c{c} or proboskís) is an iconic, highly versatile limb containing approximately 100,000 radial and lateral muscles fibers, connected to the elephant's head by an opening in the skull and controlled by a proboscis nerve \cite{Boas:1908:BP,Dagenais:2021:etal,Eales:1926,Kier:1985:KS}. Due to its exceptional characteristics, it has attracted much interest from the research community motivated by bio-mimicking its properties and movements \cite{Kaczmarski:2024a:etal,Kaczmarski:2024b:etal,Kaczmarski:2024c:etal,Leanza:2024:etal,Trivedi:2008:etal,Zhao:2020:etal,Zhang:2023:etal}. 

\begin{figure}[ht!]
\centering\includegraphics[width=0.95\textwidth]{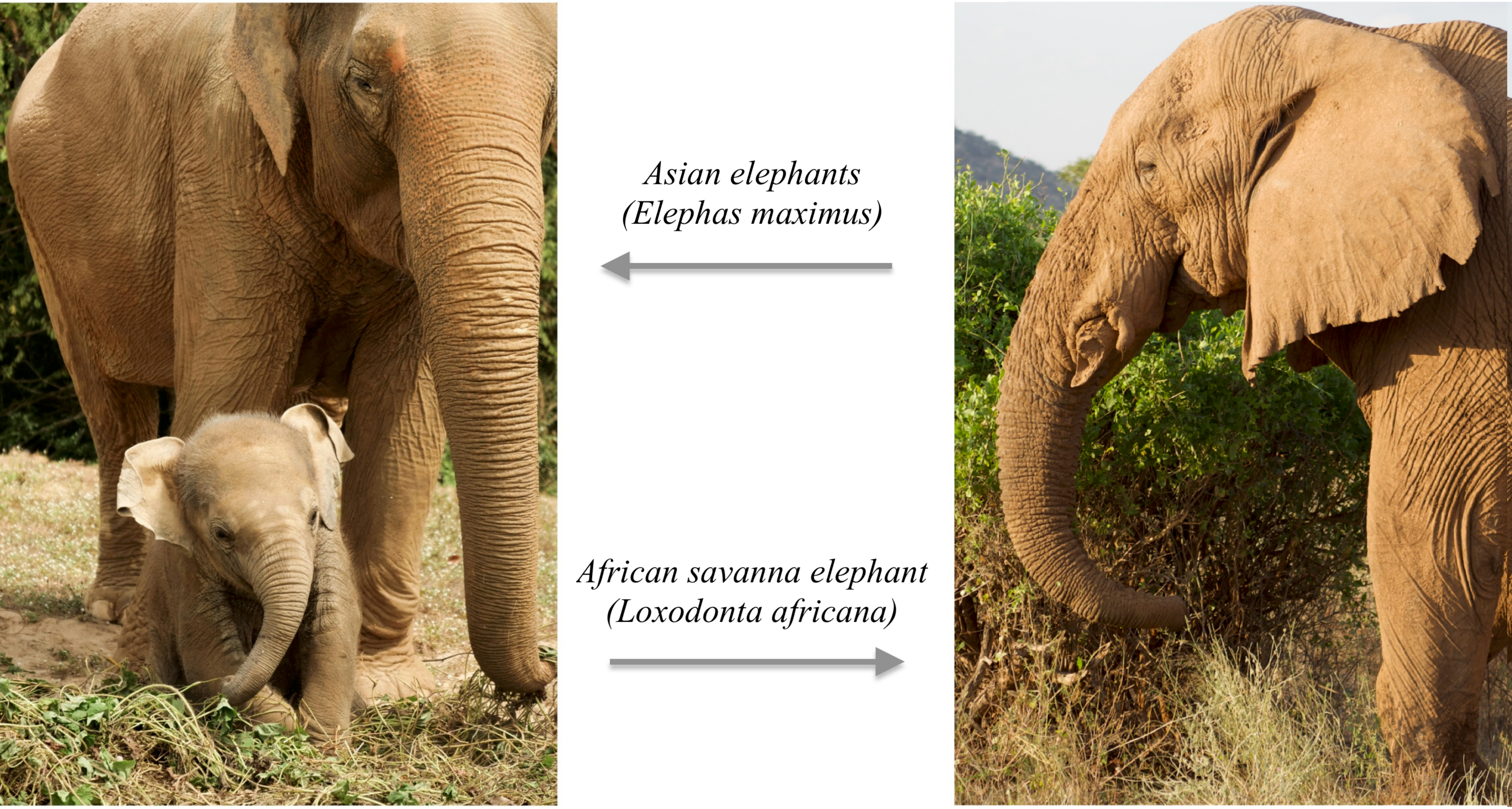}\caption{Elephants showing trunk wrinkles (photograph of African savanna elephant courtesy of Z\'ephyr Goriely).}\label{fig:elephants}
\end{figure}

A salient feature contributing to elephant trunk's agility \cite{Kaufmann:2022:etal,Longren:2023:etal,Nabavizadeh:2024,Reveyaz:2024:etal} is its rich pattern of wrinkles (see Figure~\ref{fig:elephants}), which form and develop throughout elephant's life \cite{Schulz:2020:SFSSH,Schulz:2021:SNSKBRRH,Schulz:2022:SBBSRHARHH,Schulz:2023:SRWTSMEH,Schulz:2023:SSZS,Schulz:2023:SPSBSRHH,Schulz:2024:SRKRHB}. During foetal growth, transverse trunk wrinkles and furrows appear gradually \cite{Ayer:1950:AM,Hildebrandt:2007:etal,Schulz:2024:SRKRHB}. In adult elephants, the number of transverse wrinkles differs longitudinally, with more wrinkles formed \textit{distally} (at the tip) than \textit{proximally} (near the skull), and more dorsal than ventral wrinkles. The number of wrinkles further changes with trunk-lateralization. MicroCT-imaging indicates that the outer trunk skin has a relatively constant thickness, while the inner skin is thinner within folds than between folds. Thus the generation of wrinkles in elephant trunk is primarily a result of differential growth appearing during development \cite{Thompson:1942}.

Wrinkling instability is a ubiquitous mechanism involving deformations across the scales. A general theory of small strain superposed on large strain deformations for homogeneous isotropic hyperelastic materials was developed by Green et al. \cite{Green:1952:GRS}. Particular cases of infinitesimal deformations superposed on finite extension or compression were analyzed in \cite{Fosdick:1963:FS,Wilkes1955,Woo:1962:WS}. The stability of a solid circular cylinder subject to finite extension and torsion was treated in Duka et al. \cite{Duka1993AM}. For tubes of Mooney-Rivlin material with arbitrary length and thickness, by applying the Stroh formalism \cite{Stroh:1962}, an explicit formulation for the Euler buckling load with its first nonlinear corrections was derived by Goriely et al.  \cite{Goriely2008PRSA}. A similar approach was employed in De Pascalis et al. \cite{DePascalis:2011:PDG} to calculate the nonlinear buckling load of a compressible elastic cylinder. Surface wrinkles formed by straightening of a circular sector or by bending of a cylindrical sector were examined in Destrade et al. \cite{Destrade:2014:DOSV} and Sigaeva et al. \cite{Sigaeva:2018:etal}, respectively. Wrinkling due to bending of an inflated cylindrical shell was discussed in Wu et al. \cite{Wu:2024:etal}. An extension of Fl\"{u}gge's formalism \cite{Flugge:1962,Flugge:1973} applied to the buckling of thin-walled cylinders of nonlinear elastic material was presented in Springhetti et al. \cite{Springhetti:2023:SRB}. The instability of a thick-walled hyperelastic tube subject to external pressure and axial compression was considered by Zhu et al. \cite{Zhu:2008:ZLO}. Other forms of instability, like necking or bulging, also arise during inflation of elastic tubes. These phenomena were studied extensively in \cite{Emery:2021:EF,Fu:2016:FLF,Fu:2021:FJG,Ilichev:2014:IF,Pearce:2010:PF,Ye:2020:YLF,Wang:2021:WF,Wang:2019:etal}. The influence of residual stresses on the stability of circular cylinders, and in particular on stretch-induced localized bulging and necking, was addressed by Liu and Dorfmann \cite{Liu2024MMS} and by Liu et al. \cite{LYD2024MMS}. Post-buckling modes for a core-shell cylindrical system of neo-Hookean material under axial compression with a perfectly bonded interface were investigated experimentally, theoretically and computationally in Zhao et al. \cite{Zhao2014JMPS}. The influence of geometrical incompatibility on pattern selection of growing bilayer tubes was modeled in Liu et al. \cite{Liu:2022:etal}. A cylinder with shear modulus arbitrarily varying along the radial direction which buckled and wrinkled under axial compression was examined semi-analytically by Jia et al. \cite{Jia:2014:etal}. The onset of wrinkling in an anisotropically growing fiber-reinforced tube was analyzed in Ye et al. \cite{Ye:2019:etal}. Wrinkling patterns in a cylindrical bilayer where only the outer layer grows were explored in Jia et al. \cite{Jia2015PRE}. Radially distributed wrinkles induced by a thin elastic film growing on a soft elastic substrate were presented in Jia et al. \cite{Jia:2018:etal}.

Surface wrinkling is also the primary mode of bifurcation in morphoelasticity \cite{goriely17} where growth is the bifurcation parameter for the governing system. Past the critical bifurcation point, further growth can induce various pattern transitions \cite{ZhaoJAM2017,Shen2024JMPS}. For instance, when a thin film is much stiffer than its substrate, the wrinkling mode will give way to a period-doubling pattern due to a secondary bifurcation \cite{ZhaoJAM2017,FuSIAM2015}. Moreover, if the film is around 10 times stiffer than the substrate, wrinkles can develop into period-doubling and further into folding \cite{Shen2024JMPS}. In addition, wrinkles can also directly evolve into folds or creases \cite{Stewart2016EML,Youn2024NC}. 

For elephant trunks, in \cite{Schulz:2024:SRKRHB},  major and minor wrinkles are distinguished based on the wave amplitude. This difference is similar to the period-doubling mode where one valley deepens as adjacent valleys rise \cite{FuSIAM2015}. Since these more complex patterns depend on the initial wrinkles, insights into the influence of geometric or material parameters  on wrinkling formation is critical for understanding skin patterns on  elephant trunks.

In this paper, building on the rich methodology for elastic instabilities, we construct a mathematical model that accounts for transverse skin wrinkling in elephant trunk. Working within the theoretical framework of large strain morphoelasticity, we address the following key question: \emph{What is the effect of the relative thickness, stiffness, and growth of the skin and muscle substrate on wrinkles formation along the elephant trunk?} In Section~\ref{sec:problem}, we model the trunk as a tubular cylindrical bilayer where the skin forms the thin outer layer, the muscle constitutes the thick inner layer, and a perfect bond exists between these two constituents. To account for large elastic strains, we describe each layer as an incompressible homogeneous isotropic nonlinear hyperelastic material. Assuming axisymmetric deformations, in Section~\ref{sec:numerics}, we present extensive numerical results for skin wrinkling when some model parameters change while others are fixed. In Section~\ref{sec:analytics}, we treat analytically and numerically different limiting cases. From our comprehensive analysis, we conclude that relative growth, geometry and material properties, together with loading conditions, compete to generate the characteristic pattern of transverse wrinkles in elephant trunks. While this study is motivated by a specific application scenario, our fundamental results extend further and are relevant to other applications as well.

\section{Problem formulation}\label{sec:problem}
\begin{figure}[ht!]
\centering\includegraphics[width=0.99\textwidth]{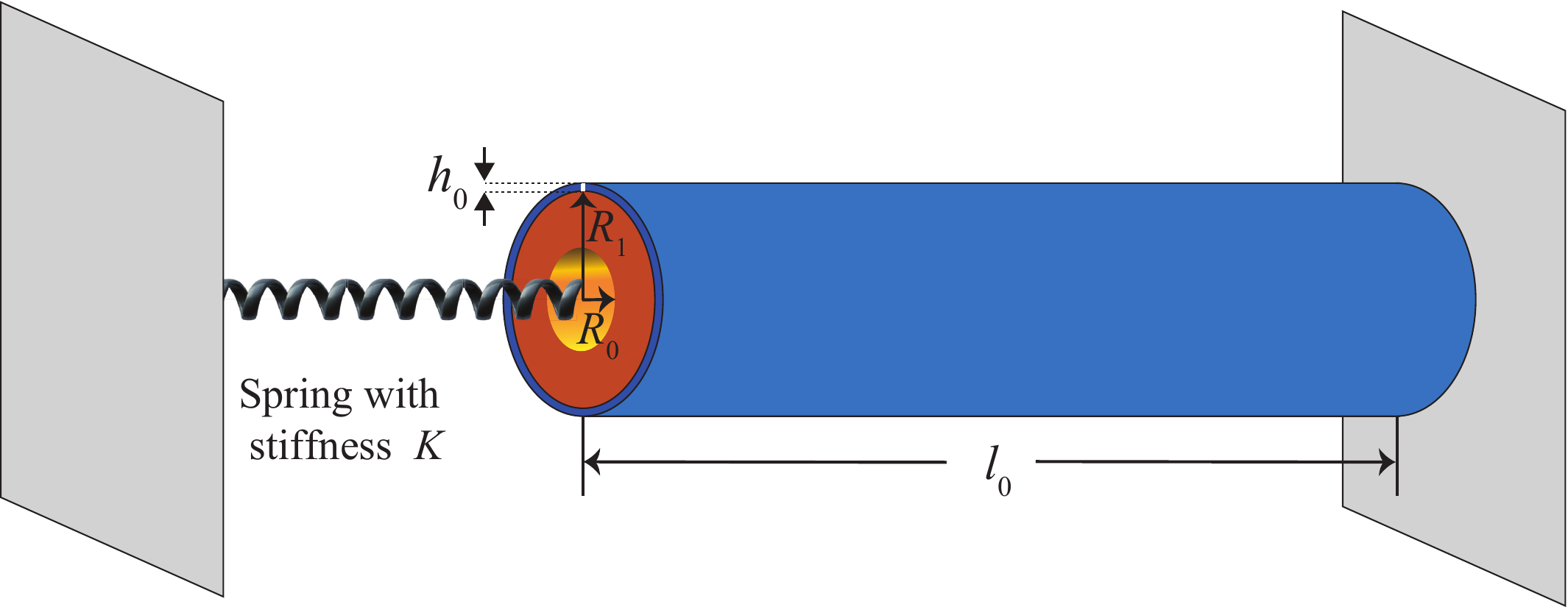}\caption{The reference bilayer cylindrical structure attached to a linear spring of stiffness $K$. The inner and outer radii are $R_0$ and $R_{1}$, respectively, the radial thickness of the shell is $h_0$, and the axial length of both the core and the shell is $l_0$.}\label{fig:tube-ref}
\end{figure}

We model (a segment of) the elephant trunk as a cylindrical tube bilayer composed of two concentric homogeneous isotropic incompressible hyperelastic tubes. The skin forms the outer shell, while the muscle substrate constitutes the inner thicker core. We denote the reference state by $\mathcal{B}_0$, and set $R_{0}$, $R_{1}$, $h_{0}$ and $l_{0}$ the inner and outer radii of the core, the uniform radial shell thickness, and the axial length of the cylindrical system, respectively, where $R_{0}<R_{1}$ and $h_{0}\ll R_{1}-R_{0}$. Assuming that both the shell and the core grow until a deformed state is attained, we designate $\mathcal{B}$ as the current configuration where the cylindrical geometry is maintained and the geometrical parameters become $r_{0}$, $r_{1}$, $h$, and $l$, respectively, with $r_{0}<r_{1}$ and $h\ll r_{1}-r_{0}$. In an elephant trunk, the skin on the dorsal side can reach $2.2$ mm in thickness, which is approximately 100 times more than typical human skin \cite{Schulz:2023:SPSBSRHH}, while the inner and outer radii are $R_0=1.5$ cm and $R_1=7.5$ cm in \cite{Schulz:2021:SNSKBRRH}, and $R_0=1.1$ cm and $R_1=2.2$ cm in \cite{Schulz:2023:SRWTSMEH}). In our mathematical model, we use these to obtain  non-dimensionalized quantities. We further assume that, at one end, the cylindrical system is fixed (homogeneous Dirichlet boundary condition), while at the other end, the system is constrained elastically (Robin boundary condition \cite{Gustafson:1998:GA}) by a linear (Hookean) spring of stiffness $K$ (Hooke's constant). Our bilayer system is represented schematically in Figure~\ref{fig:tube-ref}.

Within the usual system of cylindrical polar coordinates, we denote by $\mathbf{X}=(R,\Theta,Z)$ and $\mathbf{x}=(r,\theta,z)$ the coordinates of a material point in the reference and current configurations, respectively, where
\begin{equation}
\begin{split}
R_0\leqslant R\leqslant R_1+h_0,\qquad 0\leqslant\Theta\leqslant 2\pi,\qquad 0\leqslant Z\leqslant l_0,\\
r_0\leqslant r\leqslant r_1+h,\qquad 0\leqslant\theta\leqslant 2\pi,\qquad 0\leqslant z\leqslant l.
\end{split}
\end{equation}
For both the shell and the core, the deformation gradient $\mathbf{F}$ from the reference to the current configuration takes the form \cite{BenAmar2005JMPS,goriely17}
\begin{equation}
\mathbf{F}=\mathbf{A}\mathbf{G},\label{eq:multiplication}
\end{equation}
where $\textbf{A}$ is the elastic deformation tensor and $\textbf{G}$ is the growth tensor.

Given the strain-energy function $W(\mathbf{A})$ of an incompressible homogeneous isotropic hyperelastic material, assuming that the growth tensor $\textbf{G}$ is constant, the nominal stress tensor $\mathbf{S}$ and Cauchy stress tensor $\bm\sigma$ are, respectively,
\begin{equation}
    \mathbf{S}=J\mathbf{G}^{-1}\frac{\partial W}{\partial\mathbf{A}}-pJ\mathbf{G}^{-1}\mathbf{A}^{-1}
    \qquad\mbox{and}\qquad
    \mathbf{\bm\sigma}=\mathbf{A}\frac{\partial W}{\partial\mathbf{A}}-p\mathbf{I},
    \label{ContEqs}
\end{equation}
where $J=\det\mathbf{F}=\det\mathbf{G}$ represents the volume change due to biological growth, $p$ denotes the Lagrange multiplier associated with the incompressibility constraint for the isochoric elastic deformation 
$\det\textbf{A}=1$, and $\mathbf{I}$ is the second-order identity tensor.

In the current configuration, the equilibrium equation for a quasi-static problem without body forces takes the form
\begin{equation}
    \div\ \bm\sigma=\mathbf{0}.
\end{equation}
For the axisymmetric deformation considered here, the only non-trivial equation is
\begin{align}
\frac{\mathrm{d}\sigma_{11}}{\mathrm{d} r}+\frac{\sigma_{11}-\sigma_{22}}{r}=0,\label{eq:primary}
\end{align}
where the indices $1,2,3$, correspond to $r$-, $\theta$-, $z$-directions, respectively.
We will use  subscripts  to distinguish between physical quantities associated with the \textit{skin shell} (s) and the \textit{muscle core} (c). For example, $W_\mathrm{s}$ is the strain-energy function for the skin and $W_{\mathrm{m}}$ for the muscle. For equations that are valid for both these components, we  omit the  subscripts.

The primary deformation from $\mathcal{B}_0$ to $\mathcal{B}$ is given by
\begin{equation}
    r=r(R),\qquad \theta=\Theta,\qquad z=\lambda_zZ,\label{eq:deformation}
\end{equation}
where the longitudinal stretch ratio $\lambda_z$ is a constant. The deformation gradient then simplifies as follows:
\begin{equation}
    \mathbf{F}=\frac{\mathrm{d} r}{\mathrm{d} R}\mathbf{e}_r\otimes\mathbf{e}_r+\frac{r}{R}\mathbf{e}_\theta\otimes\mathbf{e}_\theta+\lambda_z\mathbf{e}_z\otimes\mathbf{e}_z,\label{eq:F-deformation}
\end{equation}
where $\{\mathbf{e}_{r},\mathbf{e}_{\theta},\mathbf{e}_{z}\}$ is the usual orthonormal basis for the coordinate system.

Denoting by $\{\lambda_i\}_{i=1, 2, 3}$ the principal stretch ratios in the radial, azimuthal and axial direction, respectively, by equation \eqref{eq:F-deformation}, we obtain
\begin{equation}
\lambda_1=\frac{\mathrm{d}r}{\mathrm{d}R},\qquad\lambda_2=\frac{r}{R},\qquad\lambda_3=\lambda_z.
\end{equation}

We further consider the following growth tensors for the skin and muscle, respectively:
\begin{equation}
\left\{
\begin{aligned}
&\mathbf{G}_\mathrm{s}~=~\mathbf{e}_r\otimes\mathbf{e}_r+\mathbf{e}_\theta\otimes\mathbf{e}_\theta+\gamma g\mathbf{e}_z\otimes\mathbf{e}_z,\\
&\mathbf{G}_\mathrm{m}=~\mathbf{e}_r\otimes\mathbf{e}_r+\mathbf{e}_\theta\otimes\mathbf{e}_\theta+g\mathbf{e}_z\otimes\mathbf{e}_z,
\end{aligned}
\right.
\end{equation}
where $g\geq1$ is their common growth factor and $\gamma\geq1$ is the differential (or relative) growth factor between the skin (outer layer) and the muscle (inner layer) in the axial (length) direction. Hence,
\begin{equation}
\mathbf{G}_\mathrm{s}=\mathbf{G}_\mathrm{\gamma}\mathbf{G}_\mathrm{m},
\qquad
\mbox{where}
\qquad
\mathbf{G}_\mathrm{\gamma}~=~\mathbf{e}_r\otimes\mathbf{e}_r+\mathbf{e}_\theta\otimes\mathbf{e}_\theta+\gamma \mathbf{e}_z\otimes\mathbf{e}_z.
\end{equation}
Note that, we assume homogeneous growth, with a constant (albeit different) growth factor for both the skin and the muscle. In \cite{LiuPRSA2021}, the influence of a gradient growth was found to have marginal influence on the wrinkling pattern.

From equation \eqref{eq:multiplication}, we deduce that  
\begin{equation}
\left\{
\begin{aligned}
&\mathbf{A}_\mathrm{s}~=~\alpha_{\mathrm{s}1}\mathbf{e}_r\otimes\mathbf{e}_r+\alpha_{\mathrm{s}2}\mathbf{e}_\theta\otimes\mathbf{e}_\theta+\alpha_{\mathrm{s}3}\mathbf{e}_z\otimes\mathbf{e}_z,\\
&\mathbf{A}_\mathrm{m}=~\alpha_{\mathrm{m}1}\mathbf{e}_r\otimes\mathbf{e}_r+\alpha_{\mathrm{m}1}\mathbf{e}_\theta\otimes\mathbf{e}_\theta+\alpha_{\mathrm{m}3}\mathbf{e}_z\otimes\mathbf{e}_z,
\end{aligned}
\right.
\end{equation}
and, by equation \eqref{eq:F-deformation}, we have
\begin{equation}
    \alpha_{m3}=\gamma \alpha_{s3}=\lambda_zg^{-1}.
\end{equation}

From the incompressibility conditions $\det\textbf{A}_{s}=\alpha_{s1}\alpha_{s2}\alpha_{s3}=1$ and $\det\textbf{A}_{m}=\alpha_{m1}\alpha_{m2}\alpha_{m3}$=1, together with the multiplicative decomposition \eqref{eq:multiplication}, we infer that 
\begin{equation}
r\frac{\mathrm{d}r}{\mathrm{d}R}~=~\left\{
\begin{aligned}
    &\lambda_z^{-1}\gamma gR,\quad\mbox{for}\ r_1\leqslant r\leqslant r_1+h,\\
    &\lambda_z^{-1}gR,\quad\mbox{for}\quad  r_0\leqslant r\leqslant r_1.
\end{aligned}
\right.
\end{equation}
By integration, we obtain
\begin{equation}
r^2~=~\left\{
\begin{aligned}
    &\lambda_z^{-1}\gamma g(R^2-R_1^2)+r_1^2,\quad\mbox{for}\quad r_1\leqslant r\leqslant r_1+h,\\
    &\lambda_z^{-1}g(R^2-R_0^2)+r_0^2,\hspace{6mm}\mbox{for}\quad  r_0\leqslant r\leqslant r_1.
    \label{eq:coordinate-relation}
\end{aligned}
\right.
\end{equation}
The geometry of the system in the current configuration $\mathcal{B}$ is described by
\begin{equation}\label{eq:r1hl}
\left\{\begin{aligned}
    &r_1 = \left[\lambda_z^{-1}g\left(R_1^2-R_0^2\right)+r_0^2\right]^{1/2},\\ 
    &h = \left\{\lambda_z^{-1}g\gamma\left[\left(R_1+h_0\right)^2-R_1^2\right]+r_1^2\right\}^{1/2}-r_1,\\ 
    &l = \lambda_z l_0,
\end{aligned}\right.
\end{equation}
where the deformed inner radius $r_0$ remains to be determined.

Expressing the strain-energy function in terms of the three principal elastic stretches as $W(\alpha_1,\alpha_2,\alpha_3)$, the non-zero components of Cauchy stress tensor take the following form:
\begin{equation}
\left\{
\begin{aligned}
&\sigma_{\mathrm{s}ii}~= \alpha_{\mathrm{s}i}W_{\mathrm{s}i}-p_\mathrm{s},\\
&\sigma_{\mathrm{m}ii}= \alpha_{\mathrm{m}i}W_{\mathrm{m}i}-p_\mathrm{m},
\end{aligned}
\right.
\label{eq:sigma}
\end{equation}
where $W_{\mathrm{s}i}=\partial W_\mathrm{s}/\partial\alpha_i$ and $W_{\mathrm{m}i}=\partial W_\mathrm{m}/\partial\alpha_i$, $i=1,2,3$ (no summation on repeated indices), and $p_\mathrm{s}$ and $p_\mathrm{m}$ are the Lagrange multipliers associated with the incompressibility constraint. Assuming that both the outer and inner surfaces are traction-free and, at the interface between the shell and the core, the tractions remain continuous (there is perfectly bonded contact), we have
\begin{equation}
\sigma_{\mathrm{m}11}\Big|_{r=r_0}=0,\qquad \sigma_{\mathrm{s}11}\Big|_{r=r_1+h}=0,\qquad \left(\sigma_{\mathrm{s}11}-\sigma_{\mathrm{m}11}\right)\Big|_{r=r_1}=0.
\label{eq:BC-primary}
\end{equation}

Since $\alpha_1\alpha_2\alpha_3=1$, we can define the following function depending only on two variables, 
\begin{equation}
w(\alpha,\alpha_3)=W(\alpha_1,\alpha_2,\alpha_3),
\end{equation}
where $\alpha_{2}=\alpha$ and $\alpha_{1}=\alpha^{-1}\alpha_3^{-1}$. Applying the chain rule, we obtain
\begin{equation}
w_1=\frac{\partial w}{\partial\alpha}=\alpha^{-1}(\alpha_2W_2-\alpha_1W_1),\qquad 
w_2=\frac{\partial w}{\partial\alpha_3}=\alpha_3^{-1}(\alpha_3W_3-\alpha_1W_1).
\end{equation}

From \eqref{eq:sigma}-\eqref{eq:BC-primary}, it follows that
\begin{equation}
\left\{
\begin{aligned}
&\sigma_{\mathrm{s}11}~=~\int_{\alpha_h}^{\alpha_\mathrm{s}}\frac{w_{\mathrm{s}1}}{1-\alpha^2\alpha_{\mathrm{s}3}}\mathrm{d}\alpha,\\
&\sigma_{\mathrm{m}11}=~\int_{\alpha_{r_1}}^{\alpha_\mathrm{m}}\frac{w_{\mathrm{s}1}}{1-\alpha^2\alpha_{\mathrm{m}3}}\mathrm{d}\alpha+\int_{\alpha_h}^{\alpha_{r_1}}\frac{w_{\mathrm{s}1}}{1-\alpha^2\alpha_{\mathrm{s}3}}\mathrm{d}\alpha,\\&
\hspace{3.5mm}0~~=~\int_{\alpha_{r_1}}^{\alpha_{r_0}}\frac{w_{\mathrm{s}1}}{1-\alpha^2\alpha_{\mathrm{m}3}}\mathrm{d}\alpha+\int_{\alpha_h}^{\alpha_{r_1}}\frac{w_{\mathrm{s}1}}{1-\alpha^2\alpha_{\mathrm{s}3}}\mathrm{d}\alpha,
\end{aligned}
\right.
\end{equation}
where 
\begin{equation}
    \alpha_{r_0}=\frac{r_0}{R_0},\qquad \alpha_{r_1}=\frac{r_1}{R_1},\qquad \alpha_{h}=\frac{r_1+h}{R_1+h_0}.
\end{equation}
In view of \eqref{eq:r1hl}, the above elastic stretches are connected by 
\begin{equation}\label{eq:alphar1alphah}
    \left\{\begin{aligned}
        &\alpha_{r_1}^2=\dfrac{\alpha_{r_0}^2R_0^2+g(R_1^2-R_0^2)}{R_1^2},\\
        &\alpha_h^2=\dfrac{\alpha_{r_0}^2R_0^2+g\left(R_1^2-R_0^2\right)+g\gamma\left[(R_1+h_0)^2-R_1^2\right]}{(R_1+h_0)^2}.
    \end{aligned}\right.
\end{equation}

The associated Lagrange multipliers can be determined from the following identities:
\begin{equation}
\left\{
\begin{aligned}
&p_\mathrm{s}=\alpha_{\mathrm{s}1}W_{\mathrm{s}1}-\sigma_{\mathrm{s}11},\\
&p_\mathrm{m}=\alpha_{\mathrm{m}1}W_{\mathrm{m}1}-\sigma_{\mathrm{m}11}.
\end{aligned}
\right.
\end{equation}

The resultant axial force is equal to
\begin{align}
\notag N=&~2\pi\left(\int_{r_0}^{r_1}\sigma_{\mathrm{m}33}r\mathrm{d}r+\int_{r_1}^{r_1+h}\sigma_{\mathrm{m}33}r\mathrm{d}r\right)\\
\notag=&~\pi\int_{r_0}^{r_1}\Big[2(\alpha_{\mathrm{m}3}W_{\mathrm{m}3}-\alpha_{\mathrm{m}1}W_{\mathrm{m}1})-\left(\alpha_{\mathrm{m}2}W_{\mathrm{m}2}-\alpha_{\mathrm{m}1}W_{\mathrm{m}1}\right)\Big]r\mathrm{d}r\\
&+\pi\int_{r_1}^{r_1+h}\Big[2(\alpha_{\mathrm{s}3}W_{\mathrm{s}3}-\alpha_{\mathrm{s}1}W_{\mathrm{s}1})-\left(\alpha_{\mathrm{s}2}W_{\mathrm{s}2}-\alpha_{\mathrm{s}1}W_{\mathrm{s}1}\right)\Big]r\mathrm{d}r\notag\\
=&~\pi R_0^2(\lambda_z\alpha_{r_0}^2-g)\int_{\alpha_{r_1}}^{\alpha_{r_0}}\frac{2\alpha_\mathrm{m1}w_\mathrm{m2}-\alpha w_\mathrm{m1}}{(1-\alpha^2\alpha_{\mathrm{m}3})(g-\alpha^2\alpha_{\mathrm{m}3})}\alpha\mathrm{d}\alpha\notag\\&+\pi R_1^2(\lambda_z\alpha_{r_1}^2-g\gamma)\int_{\alpha_{r_h}}^{\alpha_{r_1}}\frac{2\alpha_\mathrm{s1}w_\mathrm{s2}-\alpha w_\mathrm{s1}}{(1-\alpha^2\alpha_{\mathrm{s}3})(g\gamma-\alpha^2\alpha_{\mathrm{s}3})}\alpha\mathrm{d}\alpha.
\end{align}

Since the axial displacement is restricted at one end while the other end is attached to a spring of stiffness $K$, we can express the axial force as
\begin{equation}\label{eq:N}
    N=K(l-l_0)=K(\lambda_z-1)l_0.
\end{equation}
Then, for a given value of $g$, we can derive the axial extension $\lambda_z$ and the deformed inner radius $r_0$ from the system of equations
\begin{equation}\label{eq:N:sigmamm}
N(r_0,\lambda_z)=K(\lambda_z-1)l_0,\qquad
\sigma_\mathrm{m11}(r_0,\lambda_z)=0.
\end{equation}

For definiteness, we consider the incompressible neo-Hookean-type functions:
\begin{equation}\left\{
\begin{aligned}
    &W_\mathrm{s}=\frac{\mu_\mathrm{s}}{2}(\lambda_1^2+\lambda_2^2+\lambda_3^2-3),\\
    &W_\mathrm{m}=\frac{\mu_\mathrm{m}}{2}(\lambda_1^2+\lambda_2^2+\lambda_3^2-3),
\end{aligned}
\right.\label{eq:neo-Hookean}
\end{equation}
where $\mu_\mathrm{s}>0$ and $\mu_\mathrm{m}>0$ are the ground-state shear moduli. 

It is useful to introduce the following dimensionless quantities:
\begin{align}
&\widehat{R}_0=\frac{R_0}{l_0},\qquad 
\widehat{R}_1=\frac{R_1}{l_0},\qquad
\widehat{h}_0=\frac{h_0}{l_0},\qquad
\zeta=\frac{h_0}{R_1},\\
&\widehat{N}=\frac{N}{\pi\mu_\mathrm{m} \left[\left(R_1+h_0\right)^2-R_0^2\right]},\qquad
\widehat{\bm\sigma}=\frac{\bm\sigma}{\mu_\mathrm{m}},\qquad
\widehat{K}=\frac{K}{\pi\mu_\mathrm{m}(R_1+h_0+R_0)},\qquad
\beta=\frac{\mu_\mathrm{s}}{\mu_\mathrm{m}}.
\end{align}
Then, by equation \eqref{eq:N}, the dimensionless resultant axial force is
\begin{equation}
\widehat{N}=-\frac{\widehat{K}(\lambda_z-1)}{\widehat{R}_1+\widehat{h}_0-\widehat{R}_0}.
\end{equation}
For simplicity, henceforth, we drop the over-hat from notation. 

\begin{figure}[ht!]
	\subfigure[]{\centering\includegraphics[width=0.48\textwidth]{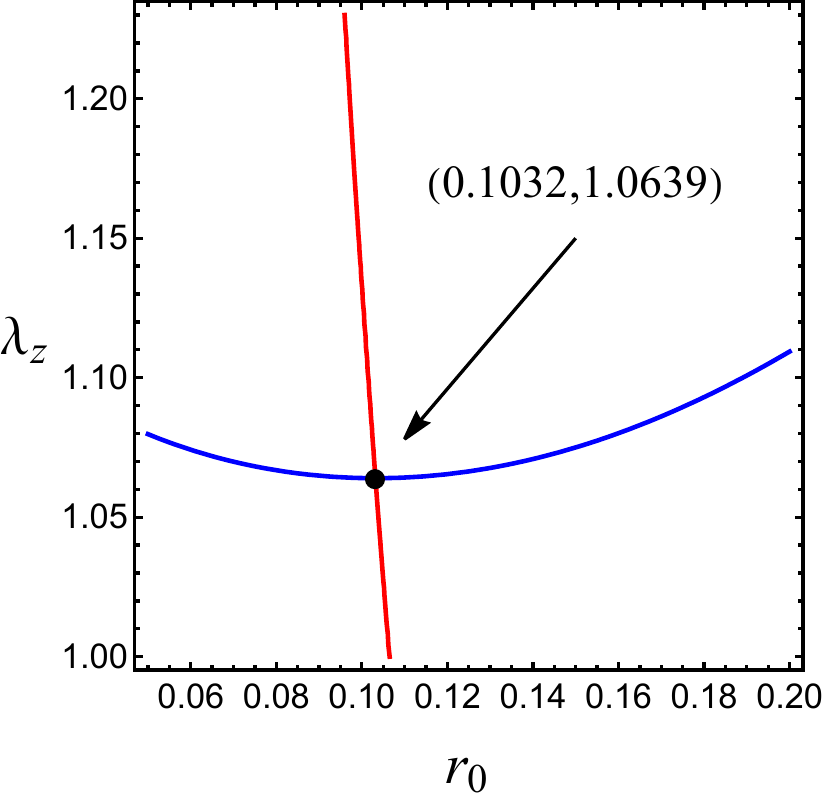}\label{fig:lambdaz-r0}}\
    \hspace{5mm}\subfigure[]{\centering\includegraphics[width=0.48\textwidth]{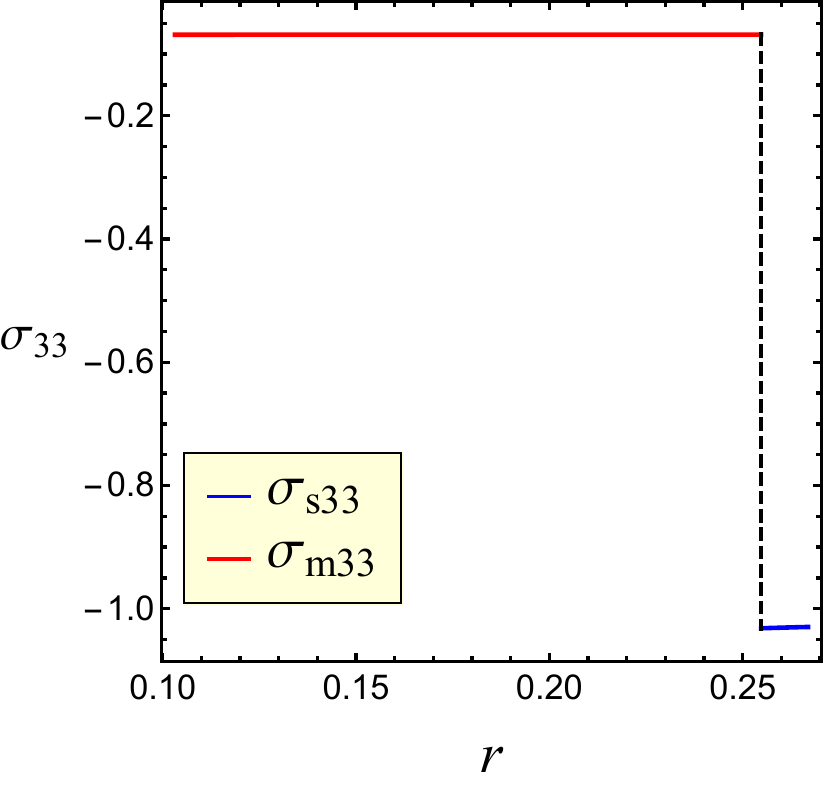}\label{fig:sigma33-r}}
    \caption{The dimensionless (a) implicit functions $\lambda_z$ and $r_0$ determined by solving the first and second equations in \eqref{eq:N:sigmamm}, and (b) axial Cauchy stress $\sigma_{33}$ as a function of the radius $r$ for the skin shell (blue) and muscle core (red) when $R_0=0.1$, $R_1=0.25$, $\zeta=0.04$ ($h_{0}=0.01$), $\beta=1.27$, $g=1.1$, $\gamma=1.2$,  and $K=1$. The dashed line in (b) indicates stress jump across the interface $r=r_1$.}\label{fig:primary-deformation}
\end{figure}

The solutions of $\lambda_z$ and $r_0$ are illustrated in Figure~\ref{fig:lambdaz-r0}, and the respective non-dimensionalized axial Cauchy stress components are indicated in Figure~\ref{fig:sigma33-r}.

Since the skin shell is stiffer than the muscle core, surface wrinkles will be generated at some critical growth. 

\section{Linear bifurcation analysis}\label{sec:numerics}

\begin{figure}[ht!]
\centering\includegraphics[width=0.7\textwidth]{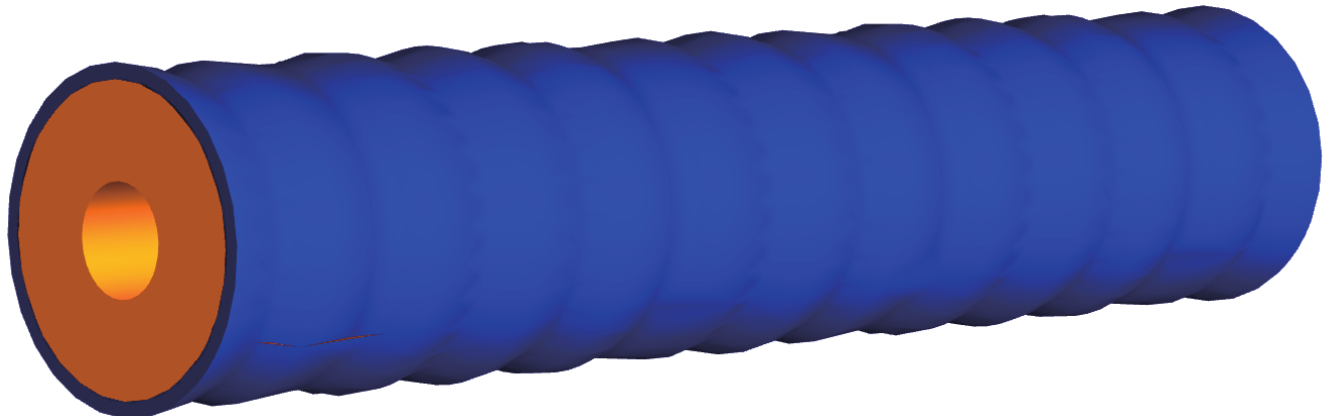}\caption{Schematic of the deformed model structure with transverse wrinkles.}\label{fig:tube-wrinkled}
\end{figure}

In this section, we derive the critical condition for transverse wrinkling (see Figure~\ref{fig:tube-wrinkled}) using the Stroh formalism \cite{Stroh:1962}. We then construct a robust numerical scheme using the surface impedance matrix method \cite{Ingerbrigtsen:1969:IT} to solve the eigenvalue problem arising from wrinkling instability. 

\subsection{Incremental theory}\label{sec:incremental-theory}
To obtain the incremental equations for the stability analysis, we denote the perturbed state for the bilayer system by $\widetilde{\mathcal{B}}$, with the associated position vector $\widetilde{\mathbf{x}}$. The relation between the position vectors in $\widetilde{\mathcal{B}}$ and $\mathcal{B}$ is
\begin{align}
   \widetilde{\mathbf{x}}(r,z)=\mathbf{x}+\bm{u},
    \label{eq:position-vector}
\end{align}
where $\bm u=u_1(r,z)\mathbf{e}_r+u_3(r,z)\mathbf{e}_z$ is the incremental displacement when $\theta$ is fixed. We focus on the axial instability \cite{Zhao2014JMPS} since major wrinkles in an elephant trunk tend to develop in the longitudinal direction \cite{Schulz:2024:SRKRHB}.

The deformation gradient from the reference configuration $\mathcal{B}_{0}$ to the perturbed configuration $\widetilde{\mathcal{B}}$ can be expressed as follows,
\begin{equation}
\widetilde{\mathbf{F}}=\frac{\partial \widetilde{\mathbf{x}}}{\partial \mathbf{X}}=(\mathbf{I}+\bm \eta)\mathbf{F},
\end{equation}
where $\bm\eta=\grad\mathbf{u}$ stands for the displacement gradient given by
\begin{equation}
    \bm \eta=\left[
    \begin{array}{ccc}u_{1,1} & u_{1,2}-\dfrac{u_2}{r} & u_{1,3} \\
    u_{2,1} & \dfrac{u_1}{r}+u_{2,2}& u_{2,3} \\
    u_{3,1} & u_{3,2} & u_{3,3}
    \end{array}
    \right]    =\left[
    \begin{array}{ccc}u_{1,1} & 0 & u_{1,3} \\
    0& \dfrac{u_1}{r}& 0 \\
    u_{3,1} & 0 & u_{3,3}
    \end{array}
    \right],
    \label{eq:displacement-gradient}
\end{equation}
with $u_{i,1}=\partial u_i/\partial r$, $u_{i,2}=\partial u_i/(r\partial\theta)$, $u_{i,3}=\partial u_i/\partial z$, $i=1,2,3$. 

For constant growth tensor $\mathbf{G}$, we have
\begin{equation}
\widetilde{\mathbf{A}}\mathbf{G}=\widetilde{\mathbf{F}}=\left(\mathbf{I}+\bm \eta\right)\mathbf{F}=\left(\mathbf{I}+\bm \eta\right)\mathbf{A}\mathbf{G},
\end{equation}
and therefore,
\begin{equation}
\widetilde{\mathbf{A}}=\left(\mathbf{I}+\bm \eta\right)\mathbf{A}.
\end{equation}

Recalling that the elastic deformation is isochoric, incompressibility implies $\det(\mathbf{I}+\bm\eta)=1$, which in its linearized form specializes to 
\begin{align}
    \operatorname{tr}\bm \eta=u_{1,1} +\frac{u_1}{r}+u_{3,3} =0,
    \label{eq:linearized-inc}
\end{align}
where `$\operatorname{tr}$' is the trace operator. 

The nominal stress tensor in $\widetilde{\mathcal{B}}$ takes the form 
\begin{align}
\widetilde{\mathbf{S}}=\widetilde{J}\mathbf{G}^{-1}\frac{\partial W}{\partial\widetilde{\mathbf{A}}}-\widetilde{p}\tilde{J}\mathbf{G}^{-1}\widetilde{\mathbf{A}}^{-1},
\end{align}
with $\widetilde{p}$ the associated Lagrange multiplier. 

We define the Lagrange multiplier increment 
\begin{equation}
\overline{p}=\widetilde{p}-p
\end{equation}
and introduce the incremental stress tensor
\begin{align}
   \bm\chi^\mathrm{T}=J^{-1}\F(\widetilde{\S}-\S)^\mathrm{T},
    \label{eq:incremental-stress}
\end{align}
where `$\mathrm{T}$' denotes the transpose operator. 

The incremental equilibrium equation reads
\begin{align}
    \operatorname{div} \bm\chi^\mathrm{T}=\mathbf{0}.
    \label{eq:incremental}
\end{align}

As the magnitude of each component of $\bm\eta$ is small, we can expand $\bm\chi$ in terms of $\bm \eta$, retaining the linear terms, i.e.,
\begin{equation}
    \chi_{ij}=\mathcal{A}_{jilk} \eta_{kl}+p\eta_{ji}-\overline{p} \delta_{ji},\quad i,j,k,l=1,2,3,
    \label{eq:expansion}
\end{equation}
where the summation convention for repeated indices is applied and $\bm {\mathcal{A}}=(\mathcal{A}_{jilk})_{i,j,k,l=1,2,3}$ denotes the first-order instantaneous modulus tensor with the following nonzero entries:
\begin{equation}
\left\{
\begin{aligned}
    &\mathcal{A}_{iijj}=\alpha_i\alpha_j W_iW_j,\\
    &\mathcal{A}_{ijij}=\frac{\alpha_iW_i-\alpha_jW_j}{\alpha_i^2-\alpha_j^2}\alpha_i^2,\quad i\neq j~\textrm{and}~\alpha_i\neq\alpha_j,\\
    &\mathcal{A}_{ijij}=\frac{\mathcal{A}_{iiii}-\mathcal{A}_{jjjj}+\alpha_iW_i}{2},\quad i\neq j~\textrm{and}~\alpha_i=\alpha_j,\\
    &\mathcal{A}_{ijji}=\mathcal{A}_{ijij}-\alpha_iW_i,\quad i\neq j.
    \label{eq:moduli-compression}
\end{aligned}
\right.
\end{equation}
Note that there is no summation for repeated indices in the above expressions and the tensor $\mathcal{A}$ has pairwise symmetry, $\mathcal{A}_{ijkl}=\mathcal{A}_{klij}$. 

By \eqref{eq:displacement-gradient}, the nonzero incremental stress components are
\begin{equation}
\left\{\begin{aligned}
&\chi_{11}=\mathcal A_{1111}\eta_{11}+\mathcal A_{1122}\eta_{22}+\mathcal A_{1133}\eta_{33}+p\ \eta_{11}-\overline{p}, \\
&\chi_{22}=\mathcal A_{2211}\eta_{11}+\mathcal A_{2222}\eta_{22}+\mathcal A_{2233}\eta_{33}+p\ \eta_{22}-\overline{p},\\
&\chi_{33}=\mathcal A_{3311}\eta_{11}+\mathcal A_{3322}\eta_{22}+\mathcal A_{3333}\eta_{33}+p\ \eta_{33}-\overline{p},\\
&\chi_{13}=\mathcal A_{3131}\eta_{13}+\mathcal A_{3113}\eta_{31}+p\  \eta_{31} ,\\
&\chi_{31}=\mathcal A_{1313}\eta_{31}+\mathcal A_{1331}\eta_{13}+p\   \eta_{13}.
\end{aligned}\right.\label{eq:components-incstr}
\end{equation}
Then equation \eqref{eq:incremental} reduces to 
\begin{equation}
   \left\{\begin{aligned}
        &\left(r\chi_{11}\right)_{,1}+r\chi_{13,3}-\chi_{22}=0,\\&
        \left(r\chi_{31}\right)_{,1}+r\chi_{33,3}=0.
    \end{aligned}\right.
    \label{inc-eq-com}
\end{equation}

We emphasize that the above general formulae are valid for both the skin and muscle layers. To specify a layer, the related subscript will be added to indicate its affiliation. 

The corresponding boundary and interface conditions are, respectively: 
\begin{equation}
 \bm\chi_\mathrm{s}\mathbf{e}_r\Big|_{r=r_1+h}=\bm 0,\qquad \left(\bm\chi_\mathrm{s}-\bm\chi_\mathrm{m}\right)\mathbf{e}_r\Big|_{r=r_1}=\bm 0
 \label{eq:bc-incremental}
\end{equation}
and
\begin{equation}
\left(u_\mathrm{s1}-u_\mathrm{m1}\right)\Big|_{r=r_1}=0,\qquad \left(u_\mathrm{s3}-u_\mathrm{m3}\right)\Big|_{r=r_1}=0.
\label{eq:incr-BC}
\end{equation}

The sliding conditions at the ends are
\begin{equation}
  \mathbf{e}_r\cdot\bm\chi_\mathrm{s}\mathbf{e}_z\Big|_{z=0,l}=0,\qquad \mathbf{e}_r\cdot\bm\chi_\mathrm{m}\mathbf{e}_z\Big|_{z=0,l}=0.\label{eq:sliding}
\end{equation}

\subsection{Stroh formalism}\label{sec:Stroh-formulation}
Employing the Stroh method \cite{Stroh:1962}, we look for periodic solutions of the form:
\begin{equation}
\left\{
    \begin{aligned}
    &u_1(r,z)=U(r)\cos kz,\qquad u_3(r,z)=W(r)\sin kz,\\&
    \chi_{11}(r,z)=T_{11}(r)\cos kz,\qquad \chi_{31}(r,z)=T_{31}(r)\sin kz,
   \end{aligned}\label{eq:periodic-solution}
\right.
\end{equation}
where $k$ represents the wave number in the axial direction. In view of equations \eqref{eq:components-incstr}, \eqref{eq:periodic-solution}, denoting the dimensionless wave number by $n$, the sliding conditions \eqref{eq:sliding} yield
\begin{equation}
    k=\frac{n\pi}{l},\quad n=1,2,3,\cdots.
\end{equation}

We further define a displacement-traction vector
\begin{equation}
    \mathbf{\Gamma}=[\mathbf{U}(r),\mathbf{T}(r)]^\mathrm{T},\quad\mbox{with}\quad \mathbf{U}(r)=[U(r),W(r)]^\mathrm{T} \quad \mbox{and}\quad \mathbf{T}(r)=[T_{11}(r),T_{31}(r)]^\mathrm{T}.
\end{equation}
From \eqref{eq:linearized-inc}, \eqref{eq:incremental}, and $\eqref{eq:components-incstr}_{1,5}$, we obtain 
\begin{equation}
    \mathbf{\Gamma}'(r)=\frac{1}{r}\mathbf{N}(r)\mathbf{\Gamma}(r),
    \label{Stroh-eq}
\end{equation}
with the prime denoting differentiation with respect to $r$ and the Stroh block matrix taking the form
\begin{equation}
    \mathbf{N}=\left[\begin{array}{cc}
        \mathbf{N}_1 & \mathbf{N}_2 \\
        \mathbf{N}_3 & -\mathbf{N}_1^\mathrm{T}
    \end{array}\right],
\end{equation}
where $\mathbf{N}_i$ $(i=1,2,3)$ are $2\times2$ sub-matrices, and, in particular, $\mathbf{N}_2$ and $\mathbf{N}_3$ are symmetric. These sub-matrices can be expressed as follows:
\begin{equation}
    \mathbf{N}_1=\left[\begin{array}{cc}
         -1& -rk \\
        \dfrac{rk(\mathcal{A}_{1331}+p)}{\mathcal{A}_{1313}} & 0
    \end{array}\right],\qquad  
    \mathbf{N}_2=\left[\begin{array}{cc}
         0& 0 \\
        0 & \dfrac{1}{\mathcal{A}_{1313}}
    \end{array}\right],\qquad 
    \mathbf{N}_3=\left[\begin{array}{cc}
        t_{11} & t_{12} \\
        t_{12} & t_{22}
    \end{array}\right],
\end{equation}
where 
\begin{equation}
\left\{\begin{aligned}
&t_{11}=\mathcal{A}_{1111}-2\mathcal{A}_{1122}+\mathcal{A}_{2222}+r^2k^2\mathcal{A}_{3131}-\frac{r^2k^2}{\mathcal{A}_{1313}}\left(\mathcal{A}_{1331}+p\right)^2+2p,\\
&t_{12}=rk\left(\mathcal{A}_{1111}-\mathcal{A}_{1122}-\mathcal{A}_{1133}+\mathcal{A}_{2233}+p\right),\\
&t_{22}=r^2k^2\left(\mathcal{A}_{1111}-2\mathcal{A}_{1133}+\mathcal{A}_{3333}+2p\right).
\end{aligned}\right.
\end{equation}

\subsection{The surface impedance matrix method}\label{sec:surface-impedance}
Next, we apply the surface impedance matrix method \cite{Biryukov1995,Shuvalov2003PRSA,Shuvalov2003QJMAM}, and introduce the conditional impedance matrix $\mathbf{Z}(r)$ through
\begin{equation}
    \mathbf{T}=\mathbf{Z}(r)\mathbf{U}.
\end{equation}
On substituting this into \eqref{Stroh-eq}, we obtain
\begin{equation}
    \mathbf{U}'=\frac{1}{r}(\mathbf{N}_1\mathbf{U}+\mathbf{N}_2\mathbf{T}),\qquad 
    \mathbf{T}'=\frac{1}{r}(\mathbf{N}_3\mathbf{U}-\mathbf{N}_1^\mathrm{T}\mathbf{T}),
\end{equation}
and the Riccati equation
\begin{equation}
    \mathbf{Z}'=\frac{1}{r}\Big(\mathbf{N}_3-\mathbf{N}_1^\mathrm{T}\mathbf{Z}-\mathbf{Z}\mathbf{N}_1-\mathbf{Z}\mathbf{N}_2\mathbf{Z}\Big).
    \label{eq:Riccati}
\end{equation}

We then apply the above general expressions to the bilayer system. For the skin layer, we use the boundary condition $\mathbf{Z}_\mathrm{m}(r_1+h)=\mathbf{0}$ determined from \eqref{eq:bc-incremental} to integrate \eqref{eq:Riccati} from $r_1+h$ to $r_1$ and find $\mathbf{Z}_\mathrm{s}(r_1)$. Applying the same procedure to the muscle layer yields $\mathbf{Z}_\mathrm{m}(r_1)$. The displacement and traction continuities at interface result in $\left(\mathbf{Z}_\mathrm{m}(r_1)-\mathbf{Z}_\mathrm{s}(r_1)\right)\mathbf{U}_\mathrm{s}(r_1)=\mathbf{0}$. For surface wrinkling, the existence of a non-trivial solution of $\mathbf{U}$ finally leads to the bifurcation condition
\begin{equation}\label{eq:bif}
\Phi(\beta,\gamma,n,g,R_0,R_1,h_0,K)=\det\left(\mathbf{Z}_{\mathrm{m}}(r_1)-\mathbf{Z}_{\mathrm{s}}(r_1)\right)=0.
\end{equation}

\subsection{Numerical examples}

We capture the interplay between different model parameters by a parametric study of the bifurcation condition \eqref{eq:bif} where some model parameters vary while others are kept fixed. For the elephant trunk, in \cite{Schulz:2022:SBBSRHARHH,Wilson:1991:etal}, the Young moduli of skin and muscle tissue are estimated as $E_{\mathrm{s}}=3\mu_{\mathrm{s}}\leq 1190\pm120$ kPa and $E_{\mathrm{m}}=3\mu_{\mathrm{m}}\approx 938$ kPa, respectively, and the skin is found at most $1.27$ times stiffer than the muscle.
Nonetheless, we should mention that those moduli were measured by deforming wrinkled skin samples, first at small strain then at large strains, and therefore, the large strain  measurements would be closer to those for the unwrinkled skin, which are not available. Additionally, these quantities will be different in the unborn elephant, where wrinkles first form \cite{Schulz:2024:SRKRHB}, than in calf and in adult elephants where wrinkles continue to evolve. Further measurements are needed. Meanwhile, our relevant plots in this paper are those where the modulus ratio is $\beta=1.27$. Notwithstanding these estimates, our modeling approach is valid more generally, and we include a wide range of examples which are valuable in their own right and can be useful to this and other applications as well.

\begin{figure}[ht!]
\subfigure[]{\centering\includegraphics[width=0.45\textwidth]{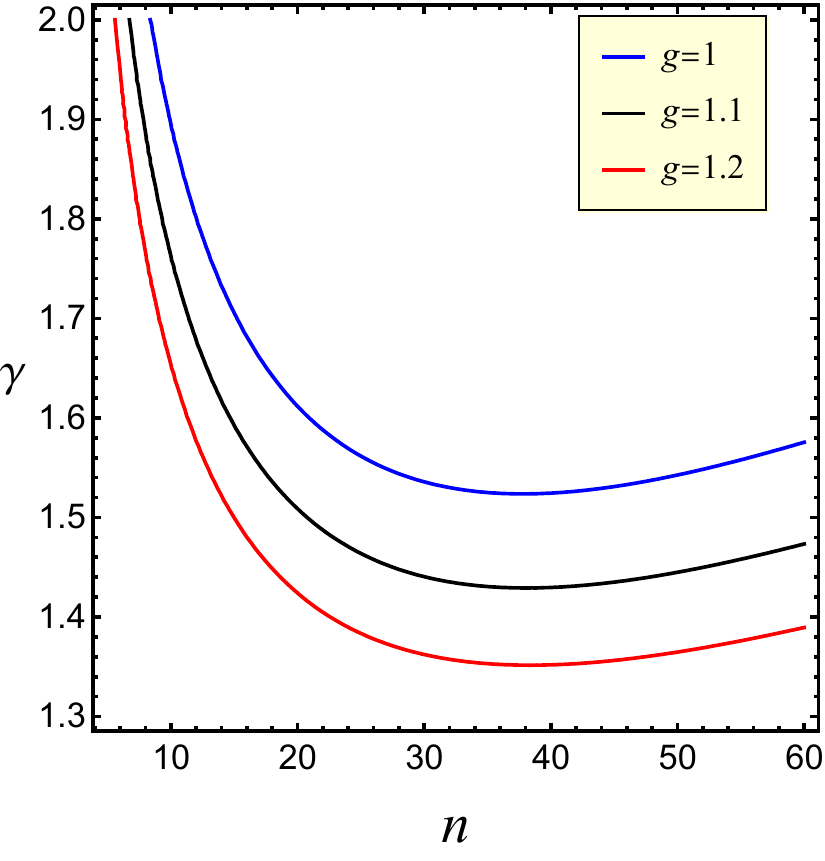}}\quad
\subfigure[]{\centering\includegraphics[width=0.455\textwidth]{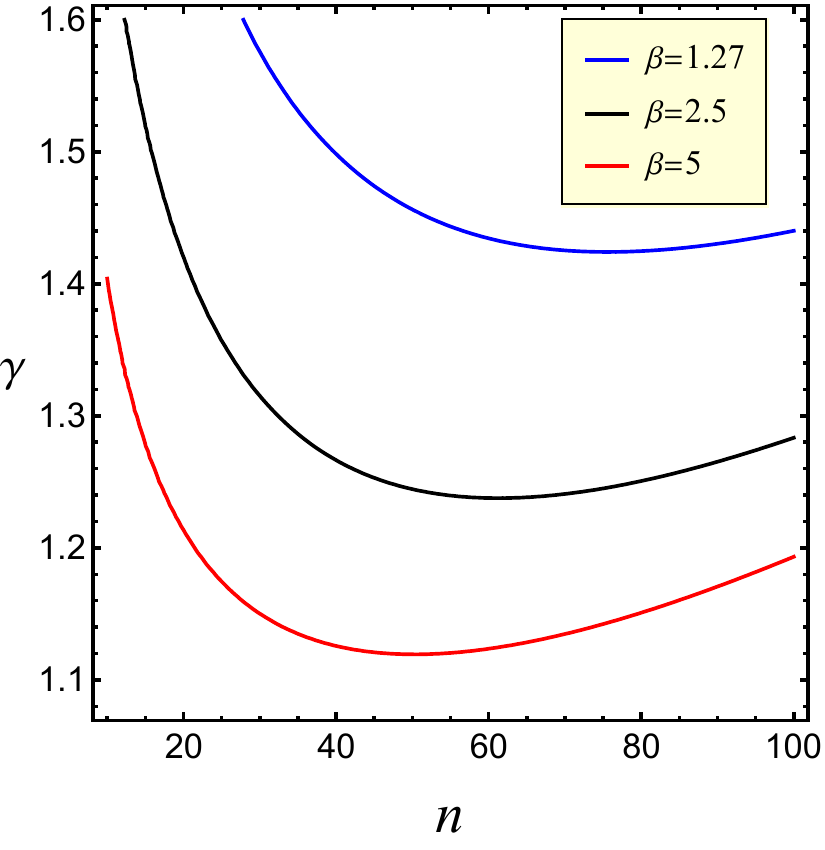}}\caption{The relative growth factor $\gamma$ as a function of the wave number $n$ identified from the bifurcation condition \eqref{eq:bif}, for different values of the pre-growth factor $g$ and of the modulus ratio $\beta$, when $R_0=0.05$, $R_1=0.25$, $\zeta=0.04$ ($h_{0}=0.01$), $K=20$, and (a) $\beta=1.27$ or (b) $g=1.1$.}\label{fig:bif-curve-1}
\end{figure}

In Figure~\ref{fig:bif-curve-1}, we present bifurcation curves identified from the solutions of \eqref{eq:bif} for different parameter values. These plots suggest that the relative growth factor $\gamma$ decreases as a function of the wave number $n$ when the pre-growth factor $g$ increases and also when the relative modulus $\beta$ increases. For each curve, the local minimum gives the first bifurcation point with the critical relative growth $\gamma_\mathrm{cr}$ and the critical wave number $n_\mathrm{cr}$. In particular, when $\beta=1.27$, the relative growth factor $\gamma\sim 1.43$ appears to be optimal.

\begin{figure}[ht!]
\subfigure[]{\centering\includegraphics[width=0.44\textwidth]{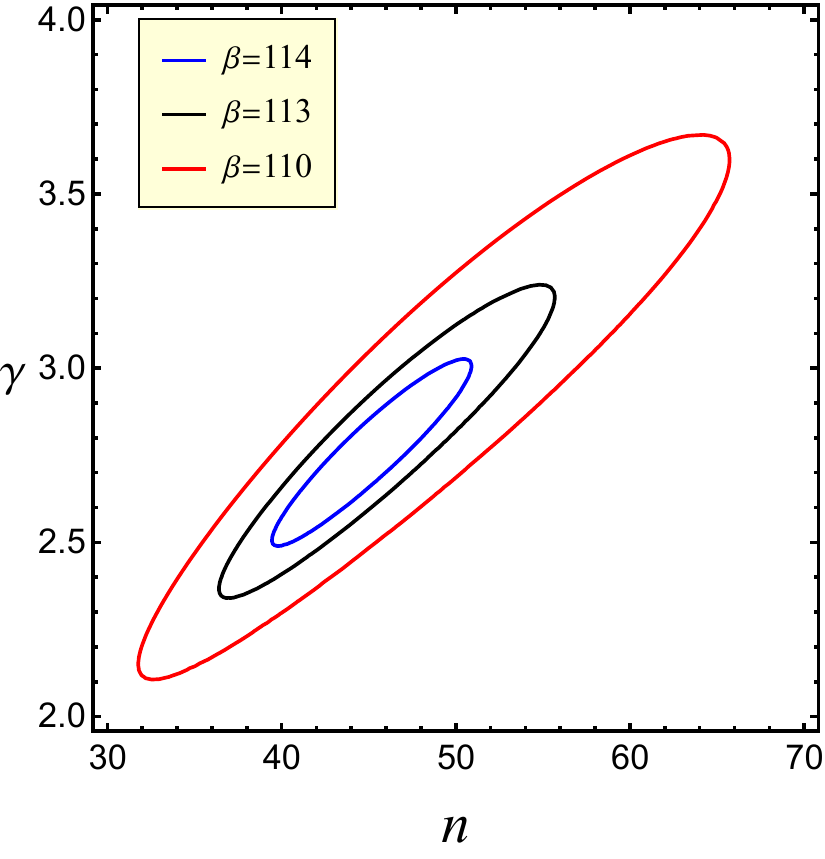}{\label{fig:gamma-n-beta65}}}\quad
\subfigure[]{\centering\includegraphics[width=0.48\textwidth]{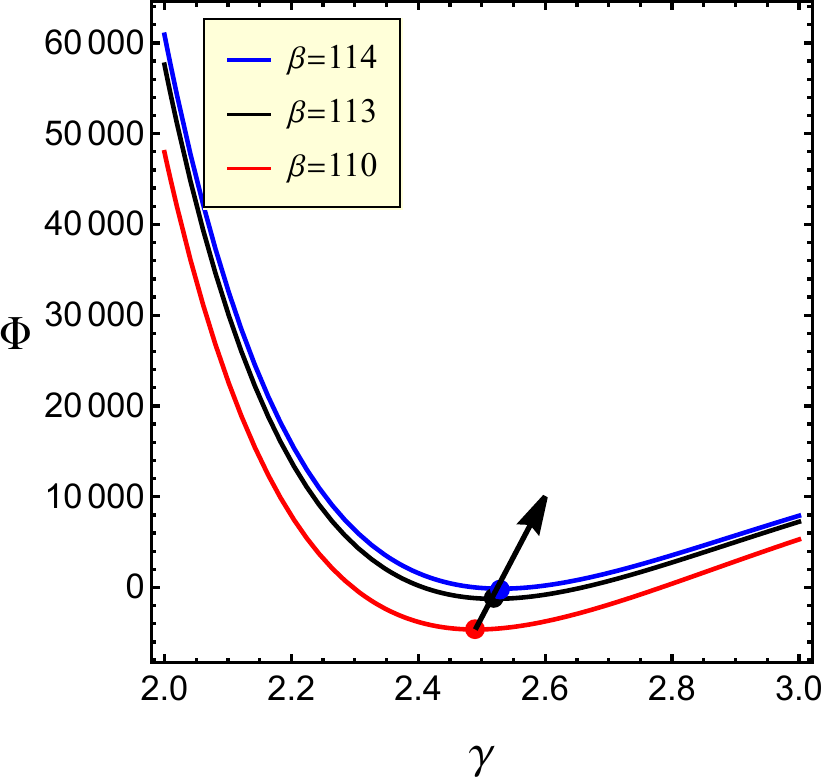}{\label{fig:phi-gamma}}}\caption{(a) The relative growth factor $\gamma$ as a function of the wave number $n$ identified from the bifurcation condition \eqref{eq:bif} when $g=1.1$, $R_0=0.1$, $R_1=0.25$, $\zeta=0.04$ ($h_{0}=0.01$), $K=0$ and (b) the function $\Phi$, given by equation \eqref{eq:bif}, depending on the growth factor $\gamma$ when the wave number is $n=40$.}\label{fig:Phi}
\end{figure}

For $\beta$ large, as illustrated in Figure~\ref{fig:gamma-n-beta65} showing bifurcation curves, all closed curves tend to shrink when $\beta$ increases. This indicates that a higher modulus ratio, corresponding to stiffer skin, delays the critical relative growth $\gamma_\mathrm{cr}$. \emph{This feature is quite different from the previously known result that a stiffer film triggers an earlier instability.} Additionally, the bifurcation condition \eqref{eq:bif} has no solution for some values of $n$.

Another inference is that, for large enough $\beta$, the bifurcation curve may vanish. To study this, we plot in Figure~\ref{fig:phi-gamma} the function $\Phi$, described by equation \eqref{eq:bif}, with the differential growth factor $\gamma$ increasing while $n=40$ and the other parameters as in Figure~\ref{fig:gamma-n-beta65}. The dots where the black arrow crosses the curves in Figure~\ref{fig:phi-gamma} highlight the minimum points, not the bifurcation points (note that the bifurcation point satisfies the bifurcation condition $\Phi=0$). The arrow indicates that  all curves increase with increasing $\beta$, until a critical value  $\beta_\mathrm{max}^{(n=40)}$ where the function $\Phi$ is tangential to the line $\Phi=0$. When $\beta$ exceeds this value,  the bifurcation condition has no solution (i.e., there is no surface wrinkling). 

In general, for  $n=n_\mathrm{cr}$, the critical value is $\beta_\mathrm{max}$, giving the global maximum value of $\beta$ for generating wrinkles. This optimal value can be identified by solving the simultaneous equations:
\begin{equation}
    \Phi=0,\qquad \frac{\partial \Phi}{\partial n}=0,\qquad \frac{\partial \Phi}{\partial \gamma}=0.
\end{equation}
The maximum modulus ratio $\beta_\mathrm{max}$ is recorded in Table \ref{tab:betamax}, for several values of the spring stiffness $K$. We find that $\beta_\mathrm{max}$ grows quickly with $K$. For instance, $\beta_\mathrm{max}\rightarrow\infty$ when $K=1$.

\begin{table}[!htb]
\centering
    \caption{Maximum modulus ratio $\beta_\mathrm{max}$, for different values of $K$, when $g=1.1$, $R_0=0.1$, $R_1=0.25$, and $\zeta=0.04$ ($h_0=0.01$).}
   	\vspace*{0.15cm}
	\scalebox{1}{
    \begin{tabular}{c||c|c|c|c|c}
    \hline
       spring stiffness $K$  & 0 & 0.02 & 0.02 & 0.03 & 0.04   \\\hline
    maximum modulus ratio $\beta_\mathrm{max}$ & 114.49 & 120.85 & 126.96 & 134.02 & 141.34\\
    \hline
    \end{tabular}
    }
    \label{tab:betamax}
\end{table}

\begin{figure}[ht!]
\subfigure[]{\centering\includegraphics[width=0.495\textwidth]{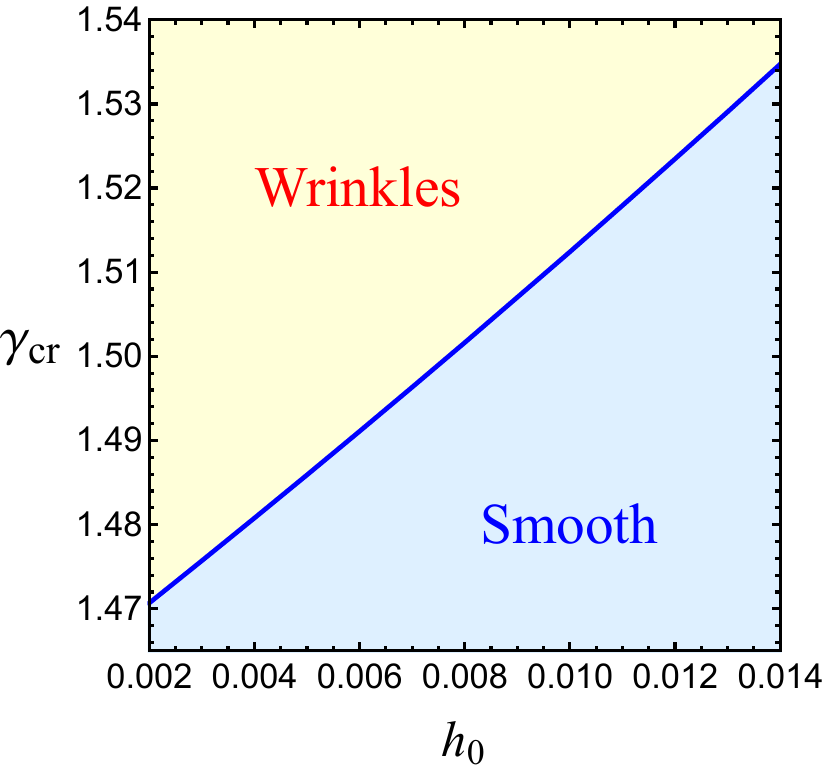}}\quad
\subfigure[]{\centering\includegraphics[width=0.48\textwidth]{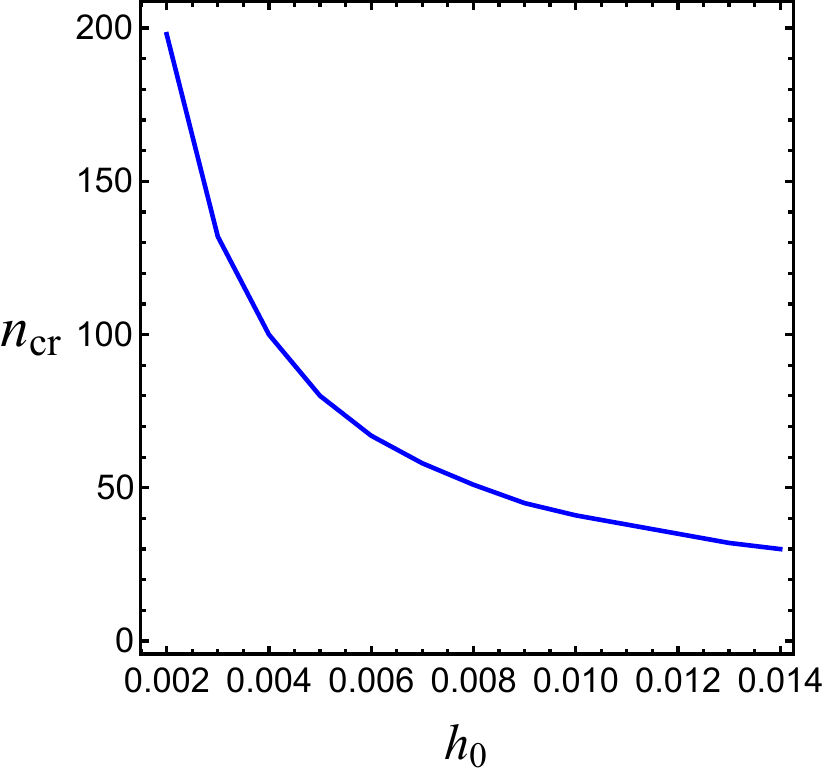}}\caption{The critical (a) relative growth factor $\gamma_\mathrm{cr}$ and (b) wave number $n_{\mathrm{cr}}$ as functions of skin thickness $h_0$ when $g=1.1$, $R_0=0.1$, $R_1=0.25$, $K=1$, and $\beta=1.27$.}\label{fig:results-h0}
\end{figure}

To investigate the effect of different parameters on the critical state, we set $\beta=1.27$. Figure~\ref{fig:results-h0} shows how the critical relative growth factor and wave number vary with respect to the skin thickness $h_0$. It suggests that a thicker skin always retards surface wrinkling and, at the same time, reduces the number of wrinkles. This is consistent with other reported results for growth-induced wrinkling \cite{goriely17}. Since $R_1=0.25$, a thicker skin layer will give rise to a lower curvature. \emph{As a result, we have that curvature increase will cause an earlier instability if the muscle thickness is fixed.}

\begin{figure}[ht!]
\subfigure[]{\centering\includegraphics[width=0.49\textwidth]{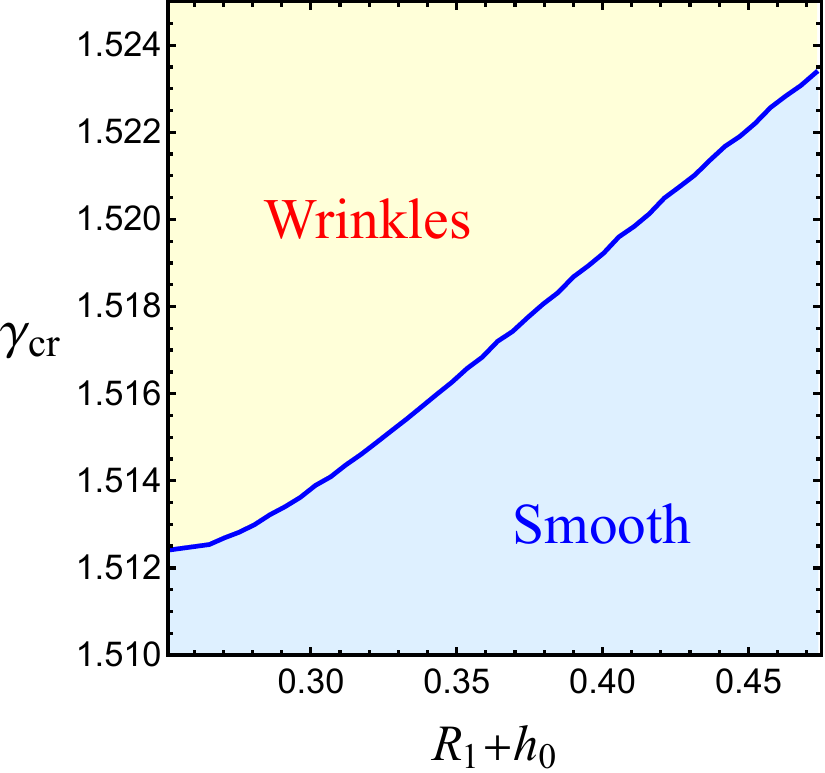}{\label{gammacr-thick}}}\quad
\subfigure[]{\centering\includegraphics[width=0.46\textwidth]{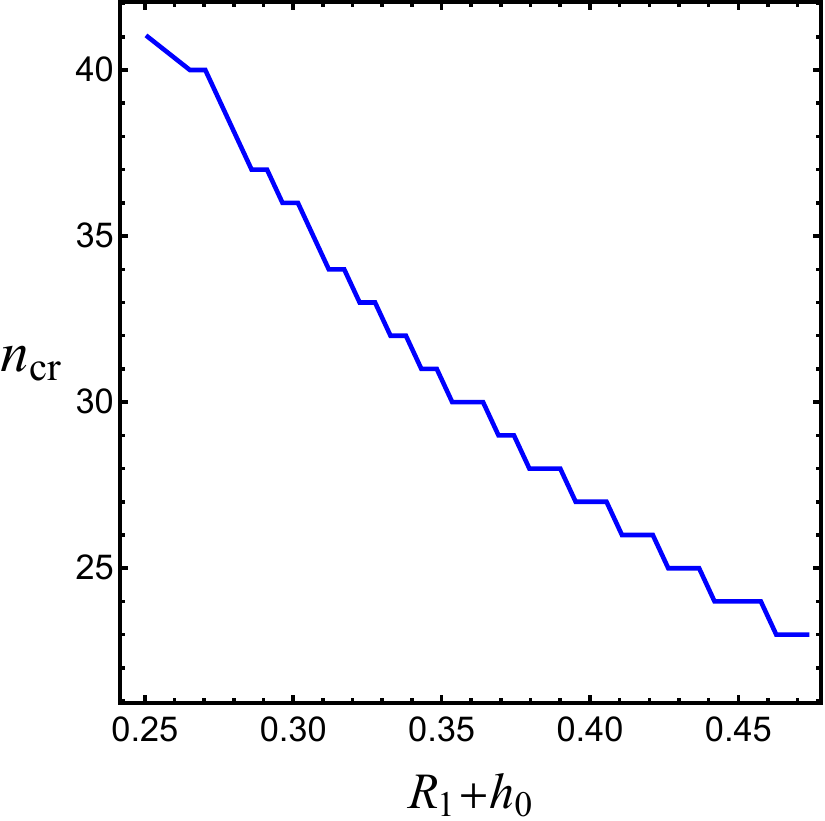}{\label{ncr-thick}}}\caption{The critical (a) relative growth factor $\gamma_\mathrm{cr}$ and (b) wave number $n_{\mathrm{cr}}$ as functions of trunk outer radius $R_{1}+h_{0}$ when $g=1.1$, $R_0=0.1$, $\zeta=0.04$, $K=1$, and $\beta=1.27$.}\label{fig:results-R1h0}
\end{figure}

The plots in Figure~\ref{fig:results-R1h0} show the dependence of $\gamma_\mathrm{cr}$ and $n_\mathrm{cr}$ on the radius of the bilayer tube $R_1+h_0$. In particular, the thickness ratio $\zeta=h_0/R_1$ is kept constant, meaning that both $h_0$ and $R_1$ increase when $R_1+h_0$ increases. It can be seen that the critical differential growth $\gamma_\mathrm{cr}$ monotonically increases, while the critical wave number $n_\mathrm{cr}$ decreases. This implies that, when the thickness ratio $\zeta$ is fixed, the critical differential growth does not alter much as the bilayer tube becomes thicker or thinner. However, the wave number changes rapidly. \emph{This may explain why there are more wrinkles in the distal narrower part of the elephant trunk than in the proximal wider part near the skull (see Figure~\ref{fig:elephants}).}

\begin{figure}[ht!]
\subfigure[]{\centering\includegraphics[width=0.495\textwidth]{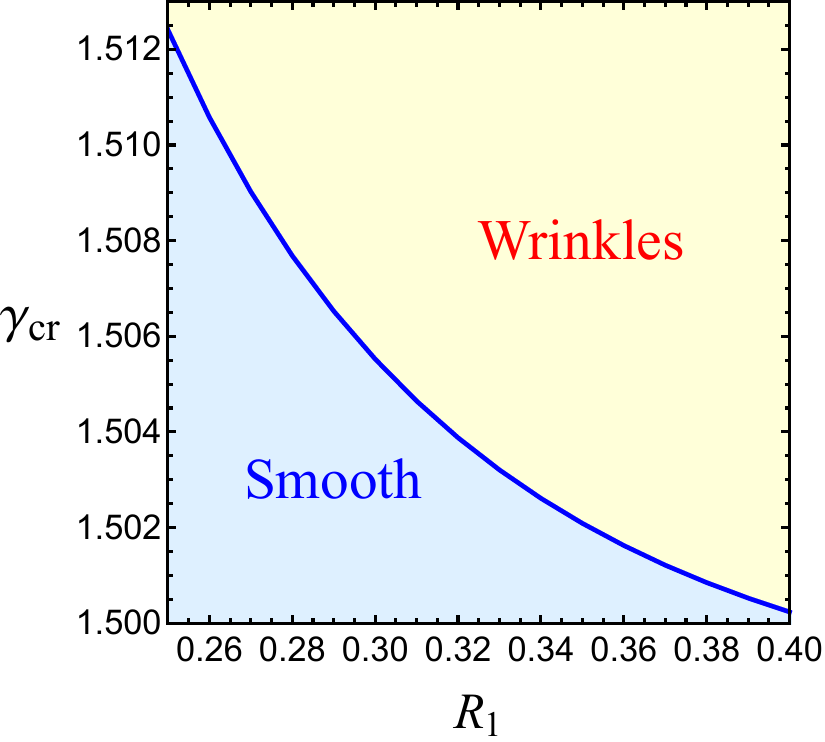}}\quad
\subfigure[]{\centering\includegraphics[width=0.46\textwidth]{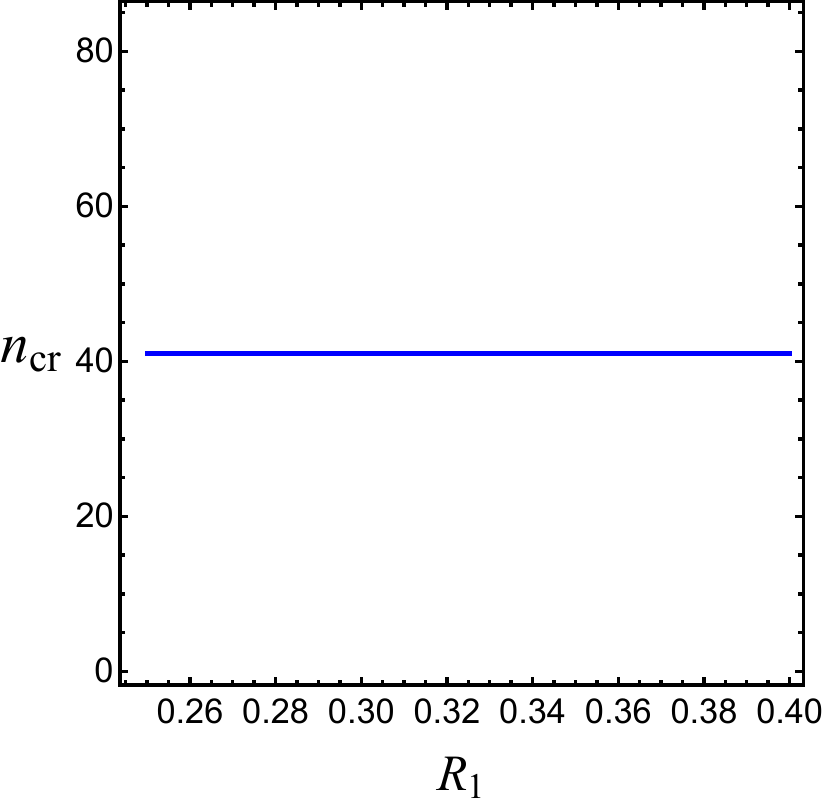}}\caption{The critical (a) relative growth factor $\gamma_\mathrm{cr}$ and (b) wave number $n_{\mathrm{cr}}$ as functions of muscle substrate outer radius $R_{1}$ when $g=1.1$, $R_0=0.1$, $\zeta=0.1$, $K=1$, and $\beta=1.27$. The critical wave number is constant.}\label{fig:results-R1}
\end{figure}

For planar film/substrate bilayers, surface wrinkles are usually not affected by the thickness of the substrate which can be seen as a half-space \cite{Liu2014IJES}. To further check whether this is true for curved structures, we display in Figure~\ref{fig:results-R1} the variations of $\gamma_\mathrm{cr}$ and $n_\mathrm{cr}$ as functions of $R_1$ by fixing all other parameters. As expected, $R_1$ alone has no effect on the wrinkled pattern but it does affect the critical differential growth $\gamma_\mathrm{cr}$. We find that a thicker muscle substrate will cause an earlier instability. Since $h_0$ is constant in this case, a higher $R_1$ corresponds to a lower curvature. \emph{We summarize by stating that curvature increase can delay surface wrinkles if the thickness of the skin layer is fixed.} Furthermore, we discover that the inner radius $R_0$ has no influence on either the critical differential growth or the critical wave number, thus we do not show the associated results here.

\begin{figure}[ht!]
\subfigure[]{\centering\includegraphics[width=0.485\textwidth]{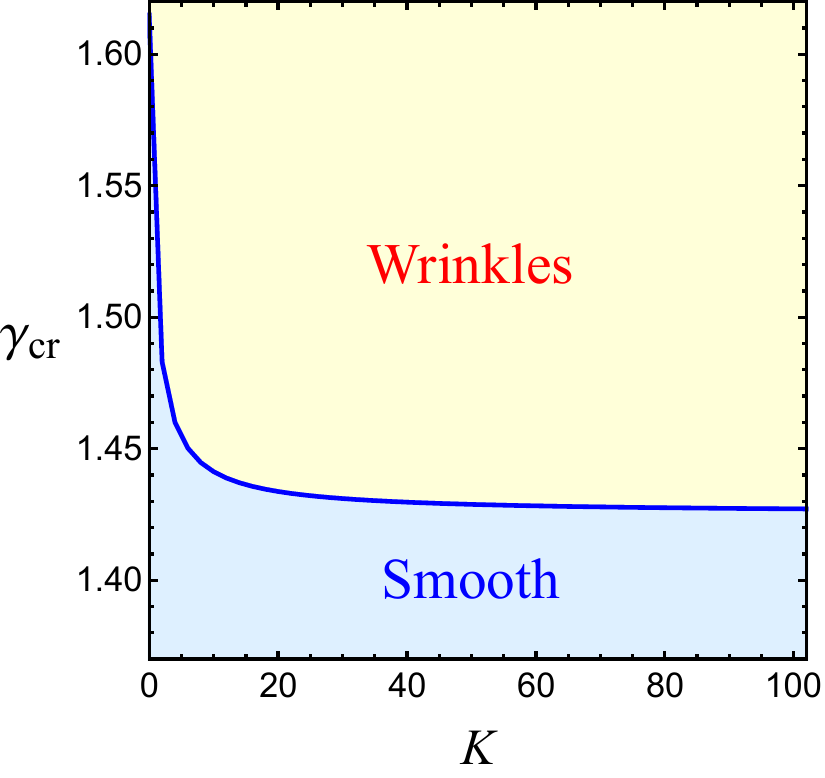}\label{gammacr-K}}\quad
\subfigure[]{\centering\includegraphics[width=0.46\textwidth]{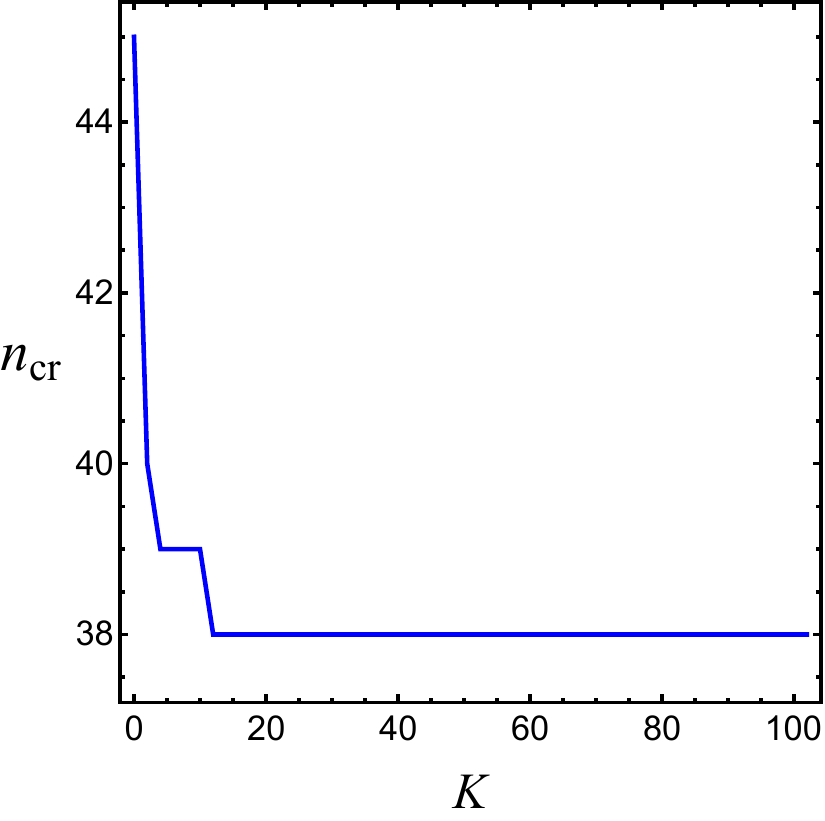}{\label{ncr-K}}}\caption{The critical (a) relative growth factor $\gamma_\mathrm{cr}$ and (b) wave number $n_{\mathrm{cr}}$ as functions of the spring stiffness $K$ at the external boundary when $g=1.1$, $R_0=0.1$, $R_1=0.25$, $\zeta=0.04$ ($h=0.01$), and $\beta=1.27$. The critical wave number remains constant and equal to $n_\mathrm{cr}=38$ if $K>10$.}\label{fig:results-K}
\end{figure}

\begin{figure}[ht!]
\subfigure[]{\centering\includegraphics[width=0.31\textwidth]{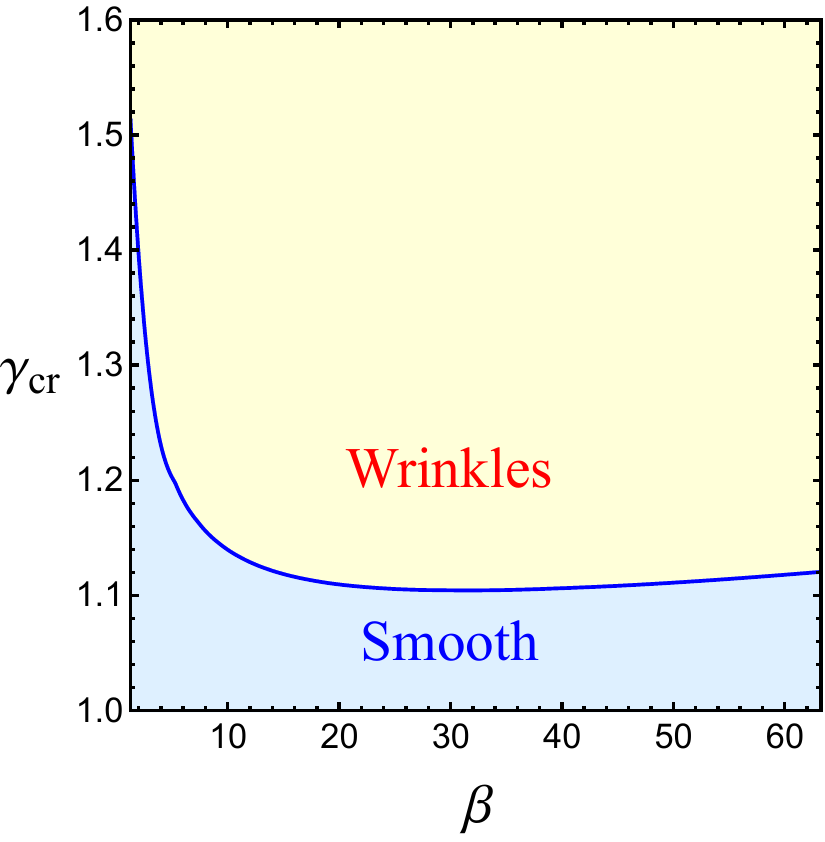}{\label{gammacr-beta}}}
\quad
\subfigure[]{\centering\includegraphics[width=0.32\textwidth]{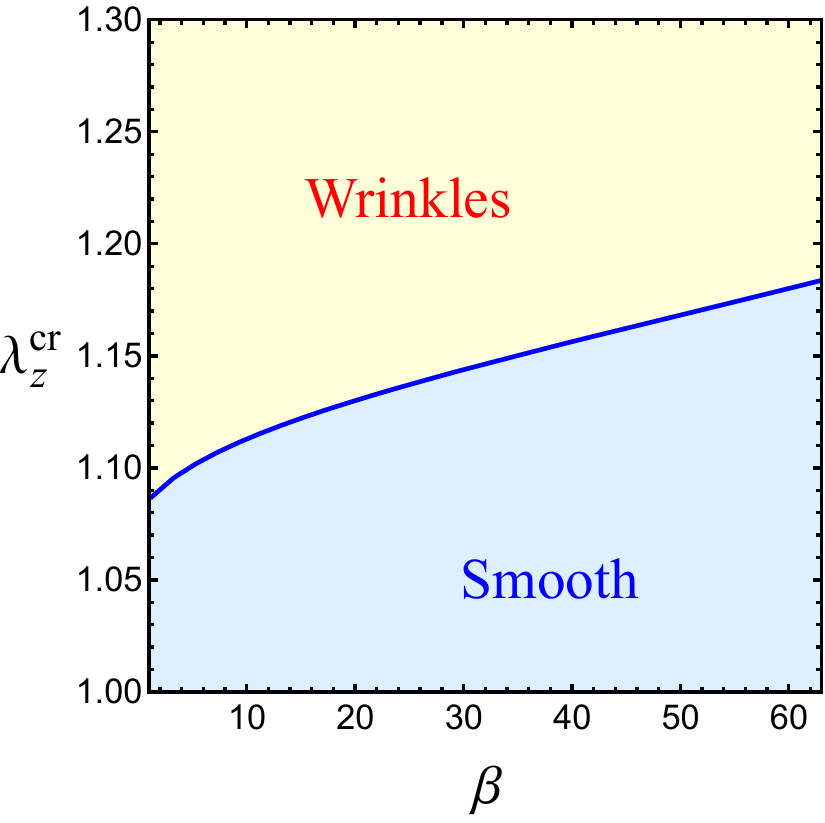}{\label{lambdacr-beta}}}
\quad
\subfigure[]{\centering\includegraphics[width=0.31\textwidth]{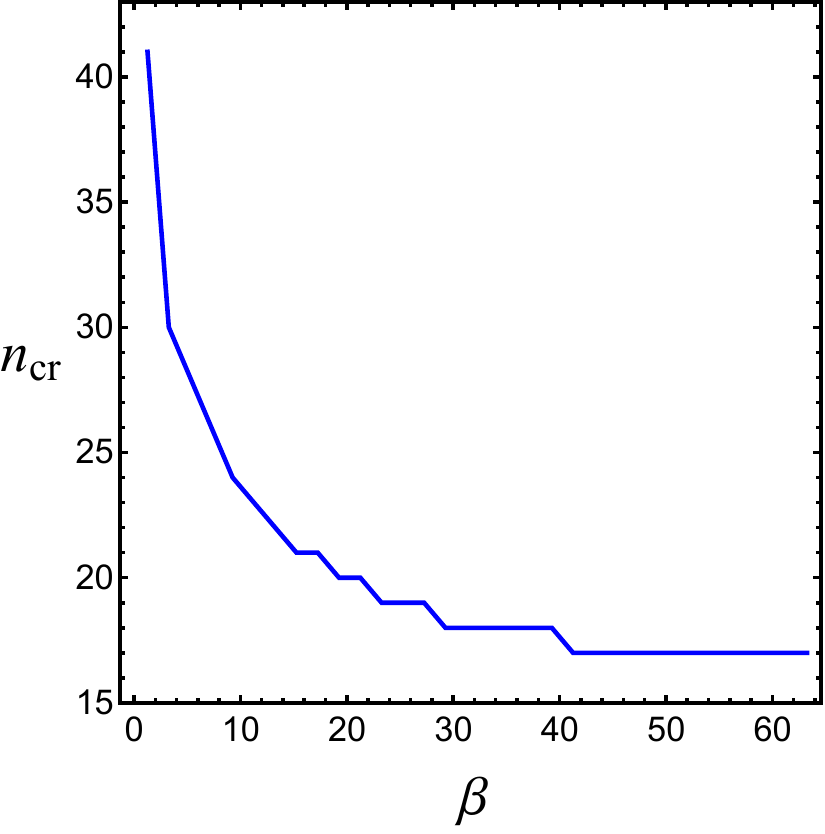}{\label{ncr-beta}}}
\caption{The critical (a) relative growth factor $\gamma_\mathrm{cr}$, (b) longitudinal stretch ratio $\lambda_{z}^{\mathrm{cr}}$ and (c) wave number $n_{\mathrm{cr}}$ as functions of the relative modulus $\beta$ when $g=1.1$, $R_0=0.1$, $R_1=0.25$, $\zeta=0.04$ ($h_0=0.01$), and $K=1$.}\label{fig:results-beta-1}
\end{figure}

Figure~\ref{fig:results-K} illustrates the effect of the spring stiffness $K$ on the critical state, namely that the critical differential growth $\gamma_\mathrm{cr}$ decreases as $K$ increases, while the critical wave number $n_{\mathrm{cr}}$ decreases from $n_\mathrm{cr}=46$ for $K=0$ (the free end without spring) to $n_\mathrm{cr}=38$ for $K\rightarrow\infty$ (the fixed length restriction).  

Figure~\ref{fig:results-beta-1} shows the critical relative growth factor $\gamma_\mathrm{cr}$ and wave number $n_{\mathrm{cr}}$ when the relative modulus $\beta$ varies. We observe from Figure~\ref{gammacr-beta} that the dependence of $\gamma_\mathrm{cr}$ on $\beta$ is non-monotonic, which is different from the case when $K\rightarrow\infty$ \cite{Jia2015PRE,Liu:2024:etal}. An analysis of the deformation reveals that it is induced by the resultant axial force $N$. By equation \eqref{eq:N}, $N$ is proportional to both $K$ and $\lambda_z$. When $\beta$ increases, corresponding to a stiffer outer layer, a larger axial force $N$ is needed to generate an instability. Figure~\ref{lambdacr-beta} presents the critical stretch $\lambda_z^\mathrm{cr}$  as a function of $\beta$. Since this function increases, more force is required to trigger surface wrinkling as $\beta$ increases. In Figure~\ref{ncr-beta}, the critical wave number $n_\mathrm{cr}$ decreases with increasing $\beta$, consistent with the other results.

\section{Asymptotic analysis}\label{sec:analytics}

To provide explicit expressions for the dependence of the critical growth rate and wave number on various parameters, we present approximate analytical results for the primary deformation and the bifurcation condition.

\subsection{Primary deformation}
To make analytical progress, we  assume $\gamma\rightarrow1$, and set $\lambda_z=1$, corresponding to the case with Dirichlet boundary conditions at both ends of the cylindrical system. This is also the limiting case when the elastic spring is infinitely stiff, i.e., $K\to\infty$. Given that the inner surface of the tubular system is free, the equation for the elastic hoop stretch $\alpha_{r_0}$ is determined by
\begin{equation}\label{eq:hoop-stretch}
    \frac{g}{\alpha_{r_0}^2}-\frac{g}{\alpha_{r_1}^2}+\beta\gamma^2\frac{g}{\alpha_{r_1}^2}-\beta\gamma^2\frac{g}{\alpha_{h}^2}+\log\frac{\alpha_{r_1}^2}{\alpha_{r_0}^2}+\beta\gamma\log\frac{\alpha_h^2}{\alpha_{r_1}^2}=0,
\end{equation}
where $\alpha_{1}^2$ and $\alpha_{h}^2$ depend on $\alpha_{r_0}^2$ through the relations \eqref{eq:alphar1alphah}. The assumption $\gamma\rightarrow1$ yields
\begin{equation}
    \alpha_{r_0}^2\sim g,\qquad \alpha_{r_1}^2\sim g,\qquad \alpha_{h}^2\sim g.
\end{equation}
We have the Taylor expansions
\begin{equation}\label{eq:alpha-r1}
   \log\frac{\alpha_{r_1}^2}{\alpha_{r_0}^2}\approx \frac{\alpha_{r_1}^2}{\alpha_{r_0}^2}-1,\qquad 
   \log\frac{\alpha_h^2}{\alpha_{r_1}^2}\approx \frac{\alpha_h^2}{\alpha_{r_1}^2}-1.
\end{equation}
Then  \eqref{eq:hoop-stretch} is approximated as follows,
\begin{equation}
    \frac{g}{\alpha_{r_0}^2}-\frac{g}{\alpha_{r_1}^2}+\beta\gamma^2\frac{g}{\alpha_{r_1}^2}-\beta\gamma^2\frac{g}{\alpha_{h}^2}+\frac{\alpha_{r_1}^2}{\alpha_{r_0}^2}-1+\beta\gamma\left(\frac{\alpha_h^2}{\alpha_{r_1}^2}-1\right)=0,
\end{equation}
or equivalently,
\begin{equation}\label{eq:hoop-stretch-approx}
\frac{\alpha_h^2}{g}\left(\frac{\alpha_{r_1}^2}{g}-\frac{\alpha_{r_0}^2}{g}\right)\left(\frac{\alpha_{r_1}^2}{g}+1\right)+\beta\gamma\frac{\alpha_{r_0}^2}{g}\left(\frac{\alpha_h^2}{g}-\frac{\alpha_{r_1}^2}{g}\right)\left(\frac{\alpha_{h}^2}{g}+\gamma\right)=0.
\end{equation}
Expressing $\alpha_{1}^2$ and $\alpha_{h}^2$ in terms of $\alpha_{r_0}^2$, as in \eqref{eq:alphar1alphah}, yields a cubic equation in $x=\alpha_{r_0}^2/g$, which admits only one real solution and can be solved directly. In particular, if $\gamma=1$, then $\alpha_{r_0}=\alpha_{r_1}=\alpha_{h}=g^{1/2}$.

Considering the scaling $R_0^2\sim h_0 l_0$, we can approximate $\alpha_{r_0}\sim g^{1/2}$ by seeking a solution of the form
\begin{equation}
    \alpha_{r_0}\approx x_0+x_1h_0+x_2h_0^2+x_3h_0^3+\cdots,
\end{equation}
where $x_i$, $i=0,1,2,\cdots$, are unknowns to be determined. After inserting the above formula into the cubic equation for $\alpha_{r_0}^2/g$, then expanding in powers of $h_0$ and collecting the coefficients of $h_0^i$, $i=0,1,2,\cdots$, we obtain infinitely many algebraic equations. Therefore we are able to solve for $x_0$, $x_1$, $x_2$ and find an asymptotic solution for $r_0=\alpha_{r_0} R_0$ of the form
\begin{equation}
    r_0\approx g^{1/2}R_0\left[1+\dfrac{\beta\gamma h_0}{2R_1}\left(1+\dfrac{8 R_0^2}{R_1^2}\right)\left(\gamma ^2-1\right)+\dfrac{3
   \beta\gamma\left(\gamma^2-1\right)^2-2\left(4\gamma^2-\gamma+3\right)\left(\gamma-1\right)}{8 R_1^2}\right].
   \label{eq:asy-r0}
\end{equation}
Note that $r_0\to g^{1/2}R_0$ as $\gamma\to 1$, which is the exact solution when there is no differential growth, i.e., the skin and the muscle grow simultaneously at the same rate. 

\begin{figure}[ht!]
\centering\includegraphics[width=0.5\textwidth]{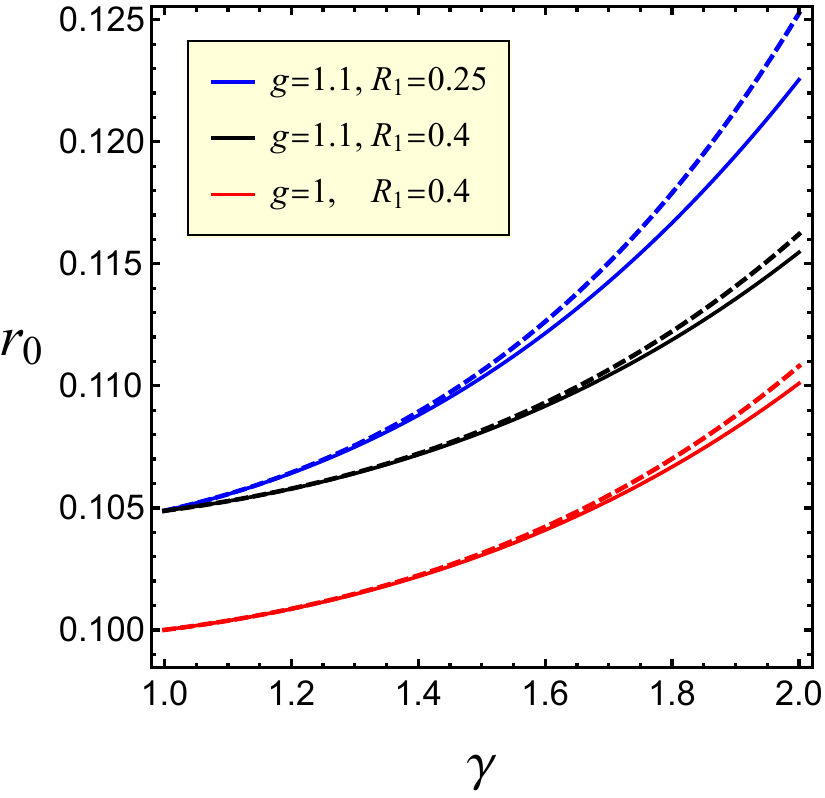}\caption{Comparisons between exact (solid) and asymptotic (dashed) solutions of $r_0$ as a function of the relative growth factor $\gamma$, for different values of $g$ and $R_1$, when $R_0=0.1$, $h_0=0.01$, and $\beta=1.27$.}\label{fig:comparison-r0-gamma}
\end{figure}

Figure~\ref{fig:comparison-r0-gamma} illustrates the exact and asymptotic solutions for $r_0$ as functions of the differential growth factor $\gamma$, when the pre-growth factor $g$ and muscle outer radius $R_1$ are given, the inner radius is $R_0=0.1$, the skin thickness is $h_0=0.01$, and the relative stiffness is $\beta=1.27$, as estimated in an elephant trunk.

\subsection{WKB approximation}

To derive analytical formulae for the critical stretch and critical wave number, we employ the WKB (Wentzel–Kramers–Brillouin) method \cite{Hinch1991}, which has been found useful in the context of wrinkling in curved and graded structures \cite{Jin:2018:etal,Fu1998IJNM,Jia:2018:etal,Liu:2024:etal}. 

Starting from the original eigenvalue problem arising from equations \eqref{inc-eq-com}, we introduce a stream function 
\begin{equation}
\psi(r,z)=f(r)\sin(kz).
\end{equation}
Then the incremental displacements are 
\begin{equation}
    u(r,z)=\frac{1}{r}\frac{\partial \psi}{\partial z},\qquad w(r,z)=-\frac{1}{r}\frac{\partial \psi}{\partial r},
\end{equation}
and the linearized incremental incompressibility condition is automatically satisfied. 

In terms of the stream function, the incremental equations and boundary conditions become, respectively,
\begin{equation}
    r^3\mathcal{A}_{1313}f''''+2r^2\left(r\mathcal{A}_{1313}'-\mathcal{A}_{1313}\right)f'''+a_2(r)f''+a_1(r)f'+a_0f=0,\label{eq:fourth-order}
\end{equation}
and
\begin{equation}
   \left\{ \begin{aligned}
        &\varrho_{\mathrm{s}11}(r_1+h)=0,\quad \varrho_{\mathrm{s}31}(r_1+h)=0,\quad\varrho_{\mathrm{m}11}(r_0)=0,\quad \varrho_{\mathrm{m}31}(r_0)=0,\\&
        \varrho_{\mathrm{s}11}(r_1)=\varrho_{\mathrm{m}11}(r_1),\quad  \varrho_{\mathrm{s}31}(r_1)=\varrho_{\mathrm{m}31}(r_1),\quad f_\mathrm{s}(r_1)=f_\mathrm{m}(r_1),\quad f^\prime_\mathrm{s}(r_1)=f^\prime_\mathrm{m}(r_1),
    \end{aligned}\right.\label{eq:bc-WKB}
\end{equation}
where
\begin{equation}
\left\{
\begin{aligned}
&\varrho_{11}=r^2\mathcal{A}_{1313}f'''+\left(r^2\mathcal{A}_{1313}'-r\mathcal{A}_{1313}\right)f''+b_1(r)f'+b_0(r)f=0,\\
&\varrho_{31}=r\mathcal{A}_{1313}f''-\mathcal{A}_{1313}f'+k^2r\left(\mathcal{A}_{1313}+p\right)f=0,
\end{aligned}
\right.
\end{equation}
with the coefficients $\{a_i\}_{i=0,1,2}$ and $\{b_i\}_{i=0,1}$ taking the following forms: 
\begin{equation}
    \left\{\begin{aligned}
        &a_0=k^2r\left(\mathcal{A}_{2222}-\mathcal{A}_{1111}+2\mathcal{A}_{1133}-2\mathcal{A}_{2233}+\mathcal{A}_{1111}'-\mathcal{A}_{1122}'-\mathcal{A}_{1331}'-\mathcal{A}_{1133}'+\mathcal{A}_{2233}'\right)\\
        &\hspace{8mm}+k^4r^3\mathcal{A}_{3131}+k^2r^3\mathcal{A}_{1331}''+k^2r^3p'',\\&
        a_1=k^2r^2\left(\mathcal{A}_{1111}-2\mathcal{A}_{1331}-2\mathcal{A}_{1133}+\mathcal{A}_{3333}\right)+k^2r^3\left(2\mathcal{A}_{1133}'-\mathcal{A}_{1111}'+2\mathcal{A}_{1331}'-\mathcal{A}_{3333}'\right)\\&
        \hspace{8mm}-3\mathcal{A}_{1313}+3r\mathcal{A}_{1313}'-r^2\mathcal{A}_{1313}'',\\&
        a_2=k^2r^3\left(2\mathcal{A}_{1133}-\mathcal{A}_{1111}+2\mathcal{A}_{1331}-\mathcal{A}_{3333}\right)+3r\mathcal{A}_{1313}-3r^2\mathcal{A}_{1313}'+r^3\mathcal{A}_{1313}',\\&
        b_0=k^2r\left(\mathcal{A}_{1111}-\mathcal{A}_{1122}-\mathcal{A}_{1133}+\mathcal{A}_{2233}+p+\mathcal{A}_{1331}'+p'\right),\\&
        b_1=k^2r\left(2\mathcal{A}_{1133}-\mathcal{A}_{1111}+\mathcal{A}_{1331}-\mathcal{A}_{3333}-p'\right)+\mathcal{A}_{1313}-r\mathcal{A}_{1313}'.
    \end{aligned}\right.
\end{equation}

For the WKB-type solution to  \eqref{eq:fourth-order}, we have 
\begin{equation}
    f=\exp\left(\int_{r_\mathrm{c}}^r\mathcal{S}(\tau)\mathrm{d}\tau\right),
    \label{eq:WKB-type}
\end{equation}
where $\mathcal{S}(r)$ is a function to be determined, the lower limit of integration is $r_\mathrm{c}=r_1$ for the skin and $r_\mathrm{c}=r_0$ for the muscle. 

To solve the eigenvalue problem with variable coefficients, arising from the mathematical model for transverse wrinkles when the wave number $n=k/\pi$ is large, we look for solutions of the form
\begin{equation}
    \mathcal{S}(r)=n \mathcal{S}_0(r)+\mathcal{S}_1(r)+\frac{1}{n}\mathcal{S}_2(r)+\cdots,
    \label{eq:WKB-expansion}
\end{equation}
where $\mathcal{S}_i$ $(i=1,2,3,\cdots)$ are unknown functions. 

Substituting the forms \eqref{eq:WKB-type} and \eqref{eq:WKB-expansion} into \eqref{eq:fourth-order} and then equating the coefficient of $n$ to zero provides an algebraic equation for $\mathcal{S}_0$. Solving this equation directly yields four independent solutions. Similarly, we are able to derive $\mathcal{S}_1$ and $\mathcal{S}_2$ in a systematic manner. 

To gain insight into the effect of different geometrical and material parameters on the critical state, we make further  simplifications. First, we see from Figure~\ref{ncr-K} that the boundary spring stiffness $K$ has a relatively minor influence on the critical wave number $n_\mathrm{cr}$, especially when $K>10$. Therefore, we can assume $K\to\infty$, which is equivalent to $\lambda_z=1$ (i.e., Dirichlet boundary conditions are imposed at both ends of the tubular system). Second, we take $\gamma=1$ so there is no differential growth and the pre-growth factor $g$ is taken as the control parameter. For the neo-Hookean material and the restricted growth case considered here, namely, $\gamma=1$ and $\lambda_z=1$, we obtain the following estimates:
\begin{equation}
        \mathcal{S}_0^{(1,2)}=\pm\pi,\quad 
        \mathcal{S}_0^{(3,4)}=\pm \pi g^{-3/2},\quad
        \mathcal{S}_1^{(1,2,3,4)}=\frac{1}{2r},\quad \mathcal{S}_2^{(1,2)}=\pm\frac{3}{8\pi r^2}, \quad \mathcal{S}_2^{(3,4)}=\pm\frac{3 g^{3/2}}{8\pi r^2}.
        \label{eq:solS0}
\end{equation}
As a result, we express the general solutions as follows
\begin{equation}
\left\{
\begin{aligned}
    &f_\mathrm{m}(r)=\sum_{i=1}^4 C_i\exp\left(\int_{r_0}^r\mathcal{S}^{(i)}_\mathrm{m}(\tau)\mathrm{d}\tau\right),\quad \mbox{for }r_0\leqslant r \leqslant r_1,\\
    &f_\mathrm{s}(r)=\sum_{i=1}^4 C_{i+4}\exp\left(\int_{r_1}^r\mathcal{S}^{(i)}_\mathrm{s}(\tau)\mathrm{d}\tau\right), \quad \mbox{for }r_1\leqslant r \leqslant r_1+h.
\end{aligned}
\right.
\end{equation}
Substituting these into the boundary and interface conditions \eqref{eq:bc-WKB}, we obtain
\begin{equation}
    \mathbf{M}\mathbf{C}=\mathbf{0},
\end{equation}
where 
\begin{equation}
\mathbf{C}=\left[C_1,C_2,C_3,C_4,C_5,C_6,C_7,C_8\right]^\mathrm{T},
\end{equation}
and
\begin{equation}
\mathbf{M}=\left[
\begin{array}{cc}
\textbf{M}_{11} & \textbf{M}_{12}\\
\textbf{M}_{21} & \textbf{M}_{22}.
\end{array}\right].
\end{equation}
The sub-matrices are
\begin{equation}
\mathbf{M}_{11}=\left[\begin{array}{cccc}
    0 & 0 & 0 & 0  \\
    0 & 0 & 0 & 0  \\ 
   \mathfrak{A}_{\mathrm{m}1}(r_0) & \mathfrak{A}_{\mathrm{m}2}(r_0) & \mathfrak{A}_{\mathrm{m}3}(r_0) & \mathfrak{A}_{\mathrm{m}4}(r_0) \\ 
    \mathfrak{B}_{\mathrm{m}1}(r_0) & \mathfrak{B}_{\mathrm{m}2}(r_0) & \mathfrak{B}_{\mathrm{m}3}(r_0) & \mathfrak{B}_{\mathrm{m}4}(r_0) 
\end{array}\right],
\end{equation}
\begin{equation}
\mathbf{M}_{12}=\left[\begin{array}{cccc}
   E_{\mathrm{s}1}\mathfrak{A}_{\mathrm{s}1}(r_1+h) & E_{\mathrm{s}2}\mathfrak{A}_{\mathrm{s}2}(r_1+h) & E_{\mathrm{s}3}\mathfrak{A}_{\mathrm{s}3}(r_1+h) & E_{\mathrm{s}4}\mathfrak{A}_{\mathrm{s}4}(r_1+h) \\
   E_{\mathrm{s}1}\mathfrak{B}_{\mathrm{s}1}(r_1+h) & E_{\mathrm{s}2}\mathfrak{B}_{\mathrm{s}2}(r_1+h) & \mathfrak{B}_{\mathrm{s}3}(r_1+h) & E_{\mathrm{s}4}\mathfrak{B}_{\mathrm{s}4}(r_1+h) \\ 
     0 & 0 & 0 & 0 \\ 
     0 & 0 & 0 & 0 
     \end{array}\right],
\end{equation}
\begin{equation}
\mathbf{M}_{21}=\left[\begin{array}{cccc}
       -E_\mathrm{m1}\mathfrak{A}_{\mathrm{m}1}(r_1) & -E_\mathrm{m2}\mathfrak{A}_{\mathrm{m}2}(r_1) & -E_\mathrm{m3}\mathfrak{A}_{\mathrm{m}3}(r_1) & -E_\mathrm{m4}\mathfrak{A}_{\mathrm{m}4}(r_1)  \\ 
     -E_\mathrm{m1}\mathfrak{B}_{\mathrm{m}1}(r_1) & -E_\mathrm{m2}\mathfrak{B}_{\mathrm{m}2}(r_1) & -E_\mathrm{m3}\mathfrak{B}_{\mathrm{m}3}(r_1) & -E_\mathrm{m4}\mathfrak{B}_{\mathrm{m}4}(r_1)  \\ 
     -E_{\mathrm{m}1}  & -E_{\mathrm{m}2} & -E_{\mathrm{m}3}  & -E_{\mathrm{m}4}  \\ 
      -E_{\mathrm{m}1}\mathcal{S}_\mathrm{m}^{(1)\prime}(r_1) & -E_{\mathrm{m}2}\mathcal{S}_\mathrm{m}^{(2)\prime}(r_1) & -E_{\mathrm{m}3}\mathcal{S}_\mathrm{m}^{(3)\prime}(r_1) & -E_{\mathrm{m}4}\mathcal{S}_\mathrm{m}^{(4)\prime}(r_1) 
\end{array}\right],
\end{equation}
\begin{equation}
\mathbf{M}_{22}=\left[\begin{array}{cccc}
      \mathfrak{A}_{\mathrm{s}1}(r_1) & \mathfrak{A}_{\mathrm{s}2}(r_1) & \mathfrak{A}_{\mathrm{s}3}(r_1) & \mathfrak{A}_{\mathrm{s}4}(r_1)   \\ 
      \mathfrak{B}_{\mathrm{s}1}(r_1) & \mathfrak{B}_{\mathrm{s}2}(r_1) & \mathfrak{B}_{\mathrm{s3}}(r_1) & \mathfrak{B}_{\mathrm{s}4}(r_1)  \\ 
    1 & 1 & 1 & 1 \\ 
      \mathcal{S}_\mathrm{s}^{(1)\prime}(r_1) & \mathcal{S}_\mathrm{s}^{(2)\prime}(r_1) & \mathcal{S}_\mathrm{s}^{(3\prime}(r_1) & \mathcal{S}_\mathrm{s}^{(4)\prime}(r_1) \\ 
\end{array}\right],
\end{equation}
with the components
\begin{equation}
\left.
\begin{aligned}
&E_{\mathrm{m}i}=\exp\left(\int_{r_0}^{r_1}\mathcal{S}_\mathrm{m}^{(i)}\mathrm{d}r\right),\quad E_{\mathrm{s}i}=\exp\left(\int_{r_1}^{r_1+h}\mathcal{S}_\mathrm{s}^{(i)}\mathrm{d}r\right),\\
&\mathfrak{A}_{i}=b_0+b_1\mathcal{S}^{(i)\prime}+r\left(r\mathcal{A}_{1313}'-\mathcal{A}_{1313}\right)\left(\mathcal{S}^{(i)\prime}\right)^2+r^2\mathcal{A}_{1313}\left(\mathcal{S}^{(i)\prime}\right)^3\\&\hspace{8mm}+r\left(r\mathcal{A}_{1313}'-\mathcal{A}_{1313}\right)\mathcal{S}_\mathrm{s}^{(i)\prime\prime}+3r^2\mathcal{A}_{1313}\mathcal{S}_\mathrm{s}^{(i)\prime}\mathcal{S}_\mathrm{s}^{(i)\prime\prime}+r^2\mathcal{A}_{1313}\mathcal{S}_\mathrm{s}^{(i)\prime\prime\prime},
\\
&\mathfrak{B}_{i}=rk^2\left(\mathcal{A}_{1313}+p\right)-\mathcal{A}_{1313}\mathcal{S}^{(i)\prime}+r\mathcal{A}_{1313}\left(\mathcal{S}^{(i)\prime}\right)^2+r\mathcal{A}_{1313}\mathcal{S}^{(i)\prime\prime},
\end{aligned}
\right\}  i=1,2,3,4.
\end{equation}
Pursuing a non-trivial solution yields
\begin{equation}
    \det\mathbf{M}=0.
\end{equation}

It can be deduced from \eqref{eq:solS0} that $E_{\mathrm{m}1}$ and $E_{\mathrm{m}3}$ are exponentially large. On the other hand, as the skin is very thin, i.e., $h$ is also small, $E_{\mathrm{s}i}$, $i=1,2,3,4$, no longer possess the exponentially large or small nature. In view of these facts, the terms proportional to $E_{\mathrm{m}1}E_{\mathrm{m}3}$ are dominant such that 
\begin{equation}
    \frac{\det\mathbf{M}}{E_{\mathrm{m}1}E_{\mathrm{m}3}}=\Psi+\mbox{exponentially small terms}.
\end{equation}
We therefore obtain an approximate bifurcation condition
\begin{equation}
    \Psi(g,n,h_0,\beta,R_0,R_1)=0.
        \label{eq-bif-explicit}
\end{equation}

\begin{figure}[ht!]
\centering\includegraphics[width=0.5\textwidth]{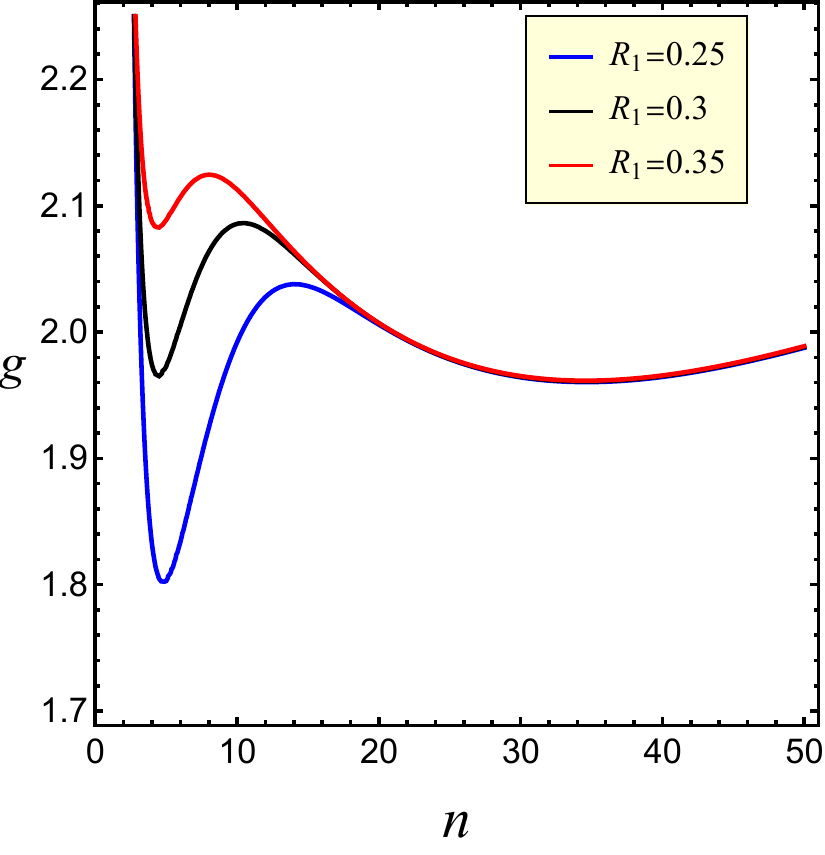}\caption{Exact bifurcation curves, for different values of the radius $R_1$ when $R_0=0.1$, $h_0=0.01$, $\beta=1.27$, and $\gamma=1$.}\label{fig:mode-transition}
\end{figure}

Before proceeding further, we specify the parameter values to be used in the subsequent calculations. To this end, we plot in Figure~\ref{fig:mode-transition} the exact bifurcation curves, in the absence of relative growth (i.e., when $\gamma=1$), for three representative values of the radius $R_{1}$ when $R_0=0.1$, $h_0=0.01$, and $\beta=1.27$. We find that each curve has two local minima: a first one which increases as $R_1$ increases, with its critical wave number decreasing, and a second one which is independent of $R_1$, and thus remains the same for all the curves. For example, when $R_1=0.25$, the  first local minimum is the global minimum, attained at the wave number $n=5$, while when $R_1=0.35$, the second local minimum is the global minimum attained at $n=35$. This suggest a possible mode transition as the aspect ratio of the bilayer tube increases. Similar mode transitions have also been reported for compressed tubes \cite{Goriely2008PRSA} and film/substrate structures \cite{Liu2014IJES,Dai2014EPL}. 

For the current problem, we can determine the critical value for the geometric parameter $R_1$ where the mode transition occurs by applying the following two-step procedure: 
\begin{itemize}
    \item[(I)] First, setting $R_1=0.35$ say, we determine $g_\mathrm{min}$ and $n_\mathrm{min}$ for the global minimum of the bifurcation curve by solving simultaneously the equations
    \begin{equation}
        \Phi(g,n)=0,\qquad  
        \frac{\partial\Phi(g,n)}{\partial n}=0.   
    \end{equation}
    \item[(II)] Second, substituting $g=g_\mathrm{min}$ in the bifurcation condition $\Phi$, we identify $R_1=R_1^{(\mathrm{m})}$ and the associated small wave number $n^{(\mathrm{m})}$ by solving the following two equations:  
\begin{equation}
    \Phi(R_1,n)=0,\qquad \dfrac{\partial\Phi(R_1,n)}{\partial n}=0.
\end{equation} 
    \end{itemize}
By applying the above procedure when $R_0=0.1$, $h_0=0.01$, $\beta=1.27$ and $\gamma=1$, as seen in Figure~\ref{fig:mode-transition}, we obtain $R_1^{(\mathrm{m})}\approx 0.2989$.
    
This interesting mode transition deserves to be treated in detail in a separate study. As our attention here is focused on wrinkling, i.e., on the case where the wave number is large, we select $R_1=0.4$ in all subsequent examples and thus avoid the mode transition mentioned above. 

\begin{figure}[ht!]
\subfigure[]
{\centering\includegraphics[width=0.46\textwidth]{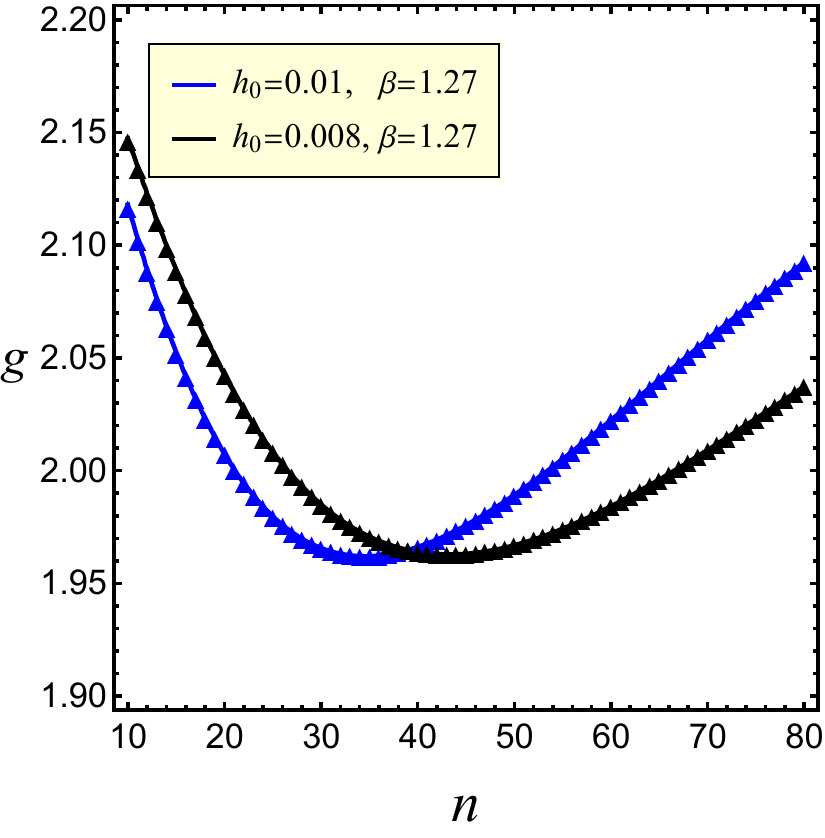}{\label{fig:g-n-comparison-beta127}}}\qquad
\subfigure[]{\centering\includegraphics[width=0.45\textwidth]{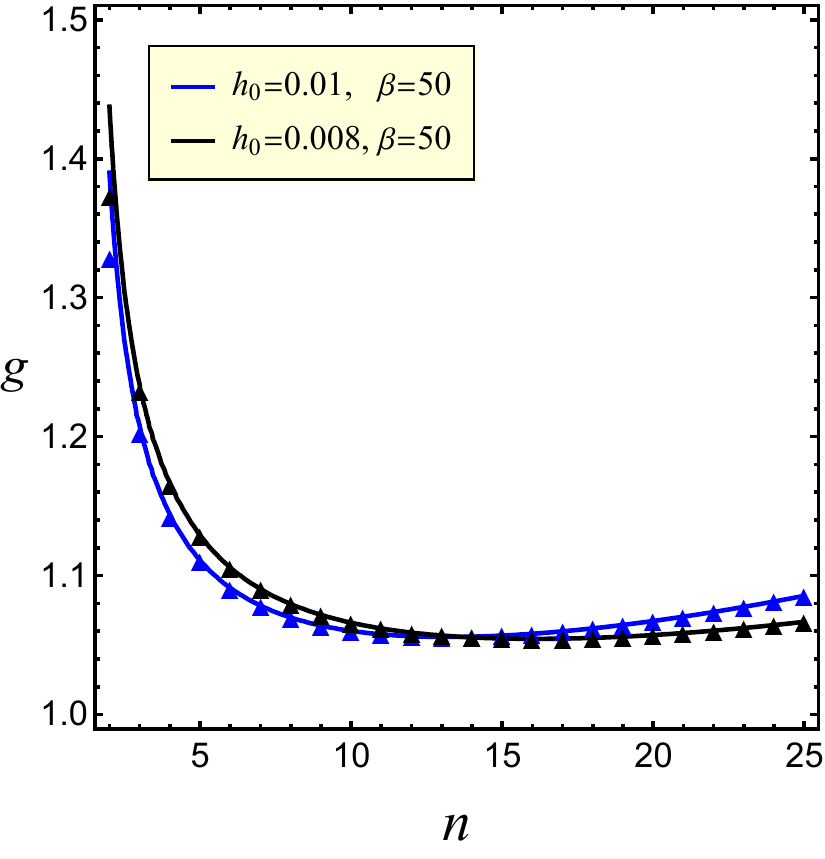}{\label{g-n-comparison}}}\caption{Comparisons between exact (solid) and asymptotic triangles) bifurcation curves for different values of $h_0$ when $R_0=0.1$, $R_1=0.4$, $\gamma=1$ and (a) $\beta=1.27$ or (b) $\beta=50$.}\label{fig:comparison-g-n}
\end{figure}

Figure~\ref{fig:comparison-g-n} compares the exact and asymptotic bifurcation curves for the pre-growth factor $g$ as a function of the wave number $n$, without differential growth (i.e., when $\gamma=1$), while the skin thickness $h_0$ is given, the muscle inner radius is $R_0=0.1$, the outer radius is $R_1=0.25$, and the modulus ratio is $\beta=1.27$ (small) or $\beta=50$ (large). As the plots for the exact and asymptotic curves completely overlap, excellent agreement between these results is obtained in our examples.

We emphasize that, when the primary deformation is homogeneous ($\gamma=1$) and the neo-Hookean model \eqref{eq:neo-Hookean} is applied,  \eqref{eq:fourth-order} can be rewritten as 
\begin{equation}
    \left(\frac{\mathrm{d}}{\mathrm{d}r^2}+\frac{1}{r}\frac{\mathrm{d}}{\mathrm{d}r}-\left(\frac{1}{r^2}+k^2\right)\right)\left(\frac{\mathrm{d}}{\mathrm{d}r^2}+\frac{1}{r}\frac{\mathrm{d}}{\mathrm{d}r}-\left(\frac{1}{r^2}+g^{-3/2}k^2\right)\right)\left(\frac{f(r)}{r}\right)=0,
    \label{eq:Bessel}
\end{equation}
which admits an analytical solution \cite{Wilkes1955,Goriely2008PRSA,Fu:2021:FJG}
\begin{equation}
    \frac{f(r)}{r}=c_1 I_1(kr)+c_2 K_1(kr)+c_3 I_1(kg^{-3/2}r)+c_4 K_1(kg^{-3/2}r),
\end{equation}
where $c_i$ ($i=1,2,3,4$) are constants, $I_1$ denotes the Modified Bessel function of the first kind and $K_1$ the Modified Bessel function of the second kind. After replacing $g$ by $\lambda_z^{-1}$ in \eqref{eq:Bessel}, we recover the scenario where a tube is subjected to end thrust, which was originally studied in \cite{Wilkes1955}. By doing so, an analytical bifurcation condition can be derived according to the boundary conditions and continuity conditions. We employ the corresponding bifurcation condition to validate the bifurcation results shown in Figures \ref{fig:mode-transition} and \ref{fig:comparison-g-n} and find exactly the same results based on \eqref{eq:bif}, despite poor computational efficiency compared to both the Stroh method and the WKB approach. However, if $\gamma\neq1$ and the primary deformation is inhomogeneous, this alternative approach becomes invalid and we need to apply WKB method.

Next, we concentrate on two distinct limiting cases, namely when $\beta\sim 1$ and when $\beta\to\infty$, respectively.

\subsubsection{The limiting case when the shear modulus ratio is close to unity} 

In an elephant trunk, the skin and muscle substrate have comparable stiffness \cite{Schulz:2022:SBBSRHARHH}, thus $\beta\sim 1$. In this case, it can be seen from Figure~\ref{fig:g-n-comparison-beta127} that the critical pre-growth factor $g_\mathrm{cr}$ is far from $1$, even when the skin layer is very thin, i.e., $0<h_0\ll 1$. In this case, we have two small parameters $h_0$ and $1/n$, and approximate the critical growth factor and critical wave number as follows \cite{Jia:2018:etal}:
\begin{equation}\label{eq:expansion-gcr}g_\mathrm{cr}=g_0+g_1h_0+g_2h_0^2+\cdots
\end{equation}
and
\begin{equation}\label{eq:expansion-ncr}        n_\mathrm{cr}=h_0^{-1}\left(n_0+n_1h_0+n_2h_0^2+\cdots\right).
\end{equation}
Inserting the above forms into the simultaneous equations given by the bifurcation condition \eqref{eq-bif-explicit} and the equation $\partial\Psi/\partial n=0$, we find that the two leading order equations for $g_0$ and $n_0$ can only be solved numerically. Once they are solved, all higher order unknowns are then derived recursively. Although we do not include here the lengthy expressions for the leading-order equations and the recursive relations, we point out that $g_0$ and $n_0$ are both independent of $R_0$ and $R_1$, and are only related to $\beta$, indicating that the critical pre-growth $g_\mathrm{cr}$ is dominated by $\beta$ only. This is useful in explaining Figure~\ref{gammacr-thick} where $\gamma_\mathrm{cr}$ belongs in the small interval $[1.5124,1.5234]$. Similarly, $n_\mathrm{cr} h_0$ is governed by $\beta$, and thus the critical wave number $n_\mathrm{cr}$ decreases as $h_0$ increases, which explains the result in Figure~\ref{ncr-thick}. 

\begin{figure}[ht!]
\subfigure[]
{\centering\includegraphics[width=0.45\textwidth]{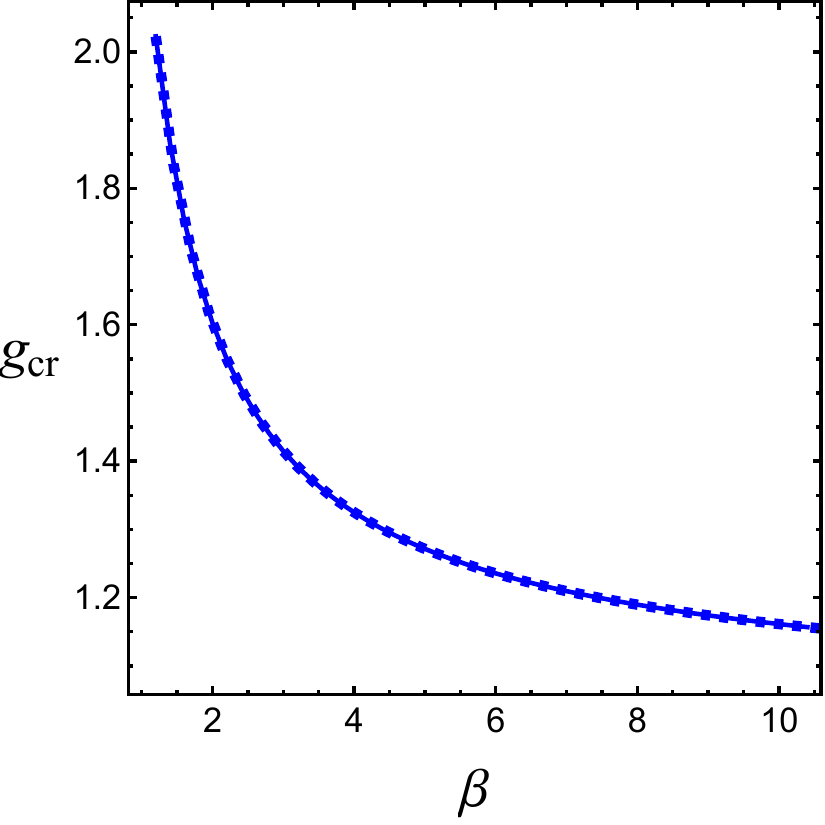}}\qquad
\subfigure[]{\centering\includegraphics[width=0.45\textwidth]{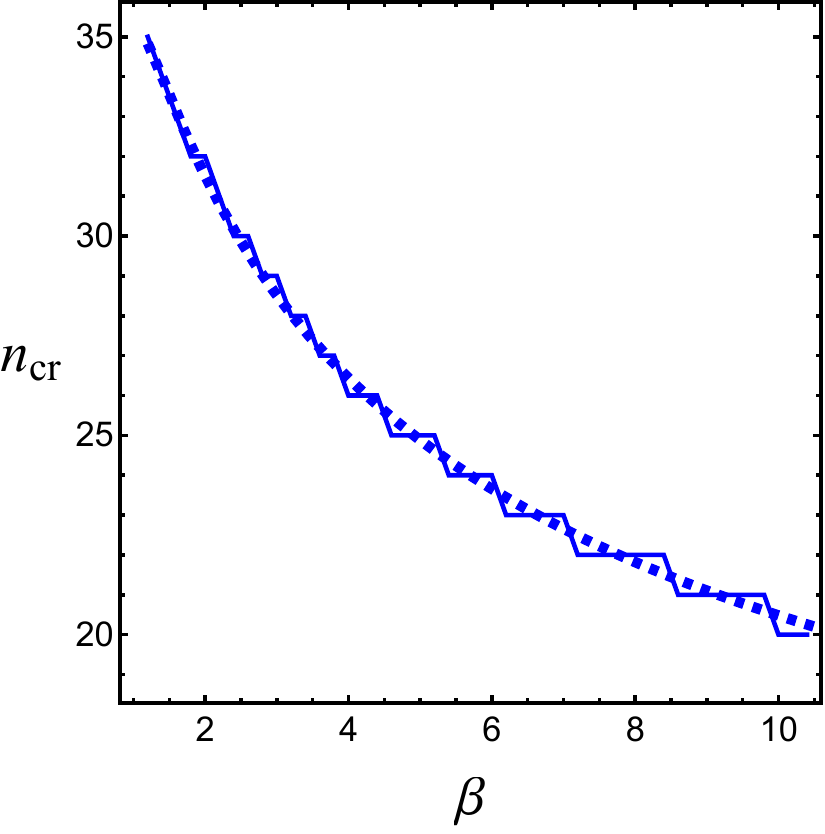}}\caption{Comparisons between exact (solid) and asymptotic (dashed) solutions for the critical pre-growth factor $g_{\mathrm{cr}}$ and critical wave number $n_{\mathrm{cr}}$ when $R_0=0.1$, $R_1=0.4$, $\zeta=0.0025$ ($h_0=0.01$), and $\gamma=1$.}\label{fig:comparison-g-n-cr-1}
\end{figure}

The three-term approximations \eqref{eq:expansion-gcr}-\eqref{eq:expansion-ncr} for the critical pre-growth factor $g_\mathrm{cr}$ and critical wave number $n_\mathrm{cr}$ when $R_0=0.1$, $R_1=0.4$, and $\zeta=0.0025$ ($h_0=0.01$) are compared with the exact solutions in Figure~\ref{fig:comparison-g-n-cr-1}. It can be seen from these plots that, for $\beta$ close to $1$, an excellent agreement is found, validating our asymptotic solutions. 

\subsubsection{The limiting case when the shear modulus ratio is large} 

When the modulus ratio $\beta$ is large, we observe from Figure~\ref{g-n-comparison} that the critical pre-growth factor is close to $1$. So there are in total four small parameters, namely, $g_\mathrm{cr}-1$, $1/\beta$, $1/n$, and $h_0$. Again, we assume that there is no differential growth, i.e., $\gamma=1$. To find analytical formulae for the critical stretch and critical wave number, we set the scales \cite{Liu2014IJES,Jin:2018:etal}
\begin{equation}
g_\mathrm{cr}-1\sim\mathcal{O}\left(\beta^{-2/3}\right),\qquad n_\mathrm{cr}\sim\mathcal{O}\left(\beta^{2/3}\right),\qquad h_0\sim\mathcal{O}\left(\beta^{-1}\right),
\end{equation}
then pursue an asymptotic solution in terms of the small parameter $1/\beta$. In a similar manner as in the previous subsection, we obtain the following explicit forms for the critical pre-growth factor 
\begin{equation}\label{eq:gcr}
\begin{split}
     g_\mathrm{cr}=1&+3^{-1/3}\beta^{-2/3}+\frac{1}{20}3^{-2/3}\beta^{-4/3}+\frac{h_0^2}{R_1^2}3^{-2/3}\beta^{2/3}-\frac{h_0}{6R_1}\beta^{-1}\\
     & +\frac{1}{2}3^{-1/3}\beta^{-5/3}+\frac{5441}{25200}\beta^{-2}
   +\frac{13 h_0^2}{20 R_1^2}-\frac{h_0^4}{3R_1^4}\beta^{2}+\mathcal{O}\left(\beta^{-7/3}\right),
\end{split}
\end{equation}
and the critical wave number
\begin{equation}\label{eq:ncr}
    \begin{aligned}
         n_\mathrm{cr}=&\frac{1}{\pi h_0}\left[3^{1/3}\beta^{-1/3}-\frac{7}{20}3^3\beta^{-1}+\frac{h_0^2}{R_1^2}\beta-\frac{h_0}+{2R_1}3^{-1/3}\beta^{-2/3}\right.\\
         &\left.\qquad \qquad+\frac{127 h_0^2}{40R_1^2}3^{-1/3}\beta^{1/3}-\frac{2 h_0^4}{R_1^4}3^{-1/3}\beta^{7/3}+\frac{47}{2800}3^{-1/3}\beta^{-5/3}+\mathcal{O}\left(\beta^{-1}\right)\right].
    \end{aligned}
\end{equation}

\begin{figure}[ht!]
	\subfigure[]
	{\centering\includegraphics[width=0.45\textwidth]{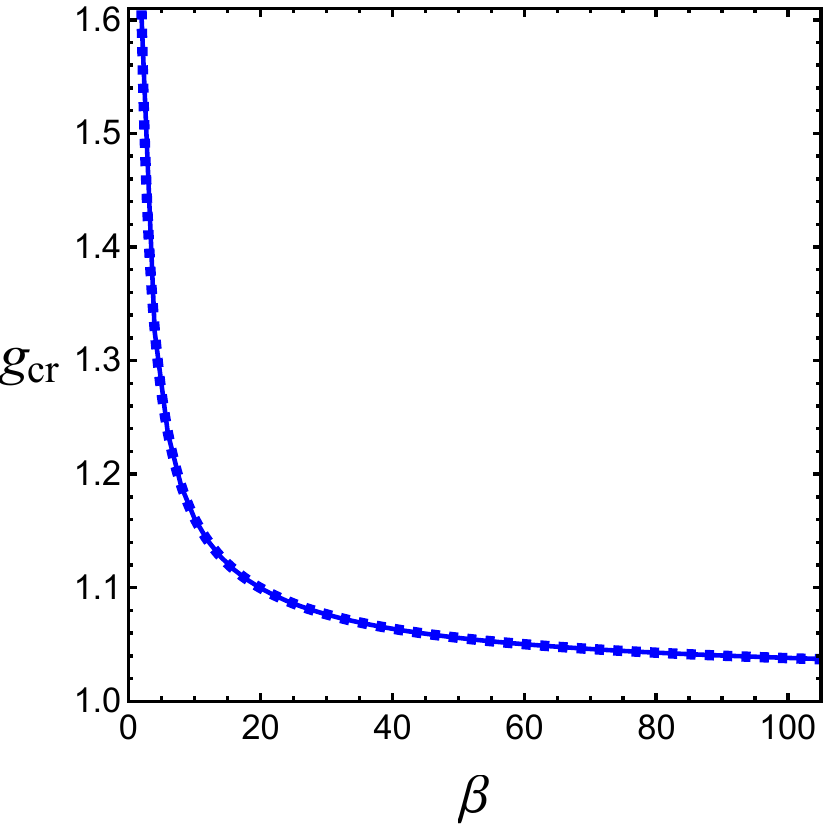}}\qquad
	\subfigure[]{\centering\includegraphics[width=0.45\textwidth]{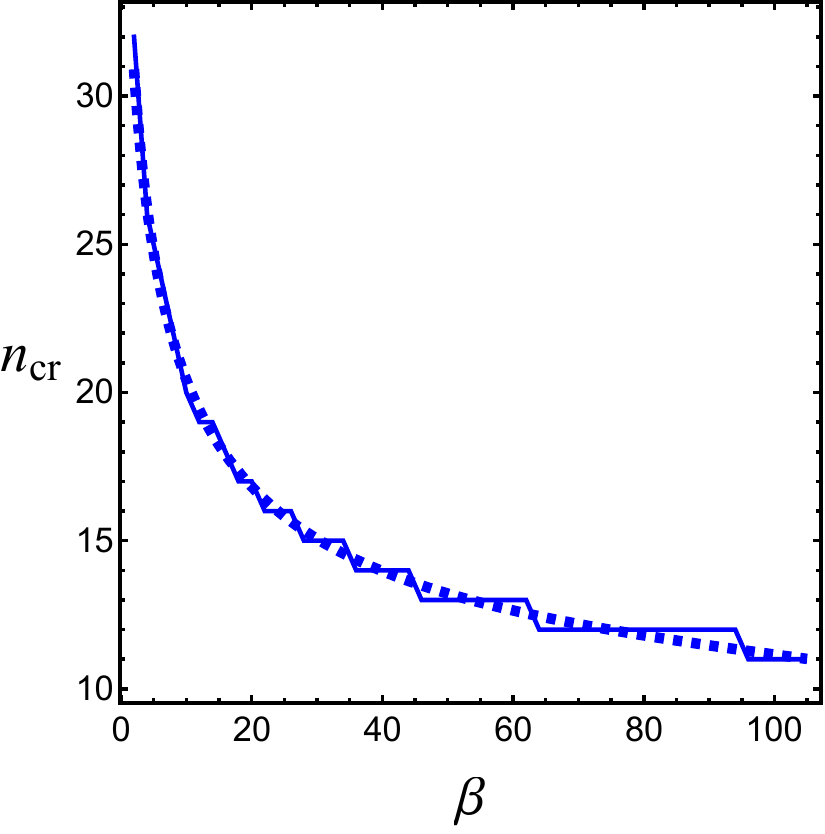}}\caption{Comparisons between exact (solid) and asymptotic (dashed) solutions for the critical pre-growth factor $g_{\mathrm{cr}}$ and critical wave number $n_{\mathrm{cr}}$ when $R_0=0.1$, $R_1=0.4$, $\zeta=0.025$ ($h_0=0.01$), and $\gamma=1$.}\label{fig:comparison-g-n-cr}
\end{figure}

From equations \eqref{eq:gcr}-\eqref{eq:ncr} we infer that the inner radius $R_0$ is not involved up to the truncated order. Meanwhile, the muscle radius $R_1$ and the skin thickness $h_0$ affect the critical pre-growth factor through higher order terms, indicating that their influence is relatively minor compared to that of the relative modulus $\beta$. In addition, both $h_0$ and $\beta$ feature in the leading-order term in  \eqref{eq:ncr}. Recalling the notation $\zeta=h_0/R_1$, we can rewrite  
\begin{equation}\label{eq:gcr-zeta}
\begin{split}
     g_\mathrm{cr}&=1+3^{-1/3}\beta^{-2/3}+\frac{1}{20}3^{-2/3}\beta^{-4/3}+\zeta^2 3^{-2/3}\beta^{2/3}-\frac{1}{6}\zeta\beta^{-1}\\
     &+\frac{1}{2}3^{-1/3}\beta^{-5/3}+\frac{5441}{25200}\beta^{-2}+\frac{13}{20}\zeta^2-\frac{1}{3}\zeta^4\beta^{2}+\mathcal{O}\left(\beta^{-7/3}\right),
\end{split}
\end{equation}
and 
\begin{equation}\label{eq:ncr-zeta}
     \begin{aligned}
         n_\mathrm{cr}&=\frac{1}{\pi h_0}\left[3^{1/3}\beta^{-1/3}-\frac{7}{20}3^3\beta^{-1}+\zeta^2\beta-\frac{\zeta}{2}3^{-1/3}\beta^{-2/3}\right.\\
         &\left.+\frac{127}{40}\zeta^2 3^{-1/3}\beta^{1/3}-2\zeta^4 3^{-1/3}\beta^{7/3}+\frac{47}{2800}3^{-1/3}\beta^{-5/3}+\mathcal{O}\left(\beta^{-1}\right)\right].
    \end{aligned}
\end{equation}
It can be seen from the above equations that $g_\mathrm{cr}$ and $n_\mathrm{cr}\pi h_0$ depend $h_0$ and $R_1$ through $\zeta$. In particular, when $\beta$ and $\zeta$ are fixed, the critical pre-growth factor $g_\mathrm{cr}$ is constant. 

Figure~\ref{fig:comparison-g-n-cr} shows the exact and asymptotic solutions for the critical pre-growth factor $g_{\mathrm{cr}}$ given by equation \eqref{eq:gcr-zeta} and the critical wave number $n_{\mathrm{cr}}$ given by equation \eqref{eq:ncr-zeta}, as functions of the modulus ratio $\beta$, when $R_0=0.1$, $R_1=0.4$, and $\zeta=0.04$ ($h_0=0.01$), in the absence of differential growth (i.e., when $\gamma=1$).  

We also remark that, in equation $\eqref{eq:ncr-zeta}$ for transverse wrinkles distributed along the tube length, the leading-order term $3^{1/3}\beta^{-1/3}$ is the same as those for the circumferential wrinkles in growing bilayer tubes 
\cite{Jin:2018:etal}, core/shell cylinders \cite{Jia:2018:etal}, and surface wrinkles in planar film/substrate bilayers \cite{Liu:2024:etal}. This analsyis further confirms the universal scaling between the modulus ratio and number of wrinkles.

\section{Conclusion}

This work stems from the need to understand wrinkling in elephant trunks where physical measurements suggest that skin and muscle substrate have comparable stiffness. Due to their mechanical compatibility, the two components can grow together seamlessly, and the wrinkled skin acts as a protective barrier that is at the same time thicker and more flexible than the unwrinkled skin. Moreover, geometric parameters, such as curvature, play key roles in the formation of transverse wrinkles. In particular, the model developed here predicts that fewer wrinkles form and earlier in the proximal region where curvature is lower than in the distal region which has larger curvature, as observed in elephant trunks. Similarly, the dorsal side presents more wrinkling than the ventral side, since curvature is higher dorsally than ventrally.

{The skin of an elephant’s trunk exhibits variations in thickness, with the ventral side differing from the dorsal side and the proximal region differing from the distal region. Despite these variations, our model remains applicable when the circular annular cross-section is adapted to represent a circular annular wedge (ventral or dorsal side) or a cylindrical segment (proximal or distal regions). Consequently, the proposed model can effectively be employed to analyze these distinct regions of the trunk individually.

Follow-up models should account for additional deformations and loading conditions, e.g., bending, unbending, causing wrinkling lateralization, and inflation under internal pressure, as in a water filled elephant trunk. The cylindrical geometry used in this study could be further refined by adopting a more realistic truncated conical shape. Additionally, more advanced hyperelastic strain-energy functions could be used to better represent the mechanical properties of skin and muscle tissues. 

The elephant trunk is a remarkable source of inspiration for bio-mechanical devices, yet its physiology remains insufficiently understood. We hope that our analysis will encourage further quantitative investigations into the mechanics of the elephant trunk and the unique properties of elephant skin. While our theoretical and numerical results describe mathematically how transverse wrinkles form in elephant trunks, our investigation is an important addition to the wrinkling methodology, and can be useful to other applications as well. 

\paragraph{Acknowledgment.}
We gratefully acknowledge the UKRI Horizon Europe Guarantee MSCA (Marie Skłodowska-Curie Actions) Postdoctoral Fellowship to Yang Liu (EPSRC EP/Y030559/1). Yang Liu further acknowledges the financial support from the National Natural Science Foundation of China (Project Nos. 12072227 and 12021002). 

\paragraph{Data availability statement.}
There are no additional data associated with this article.

\paragraph{Conflict of  interest.}
The authors have no competing interests to declare that are relevant to the content of this article.


\end{document}